\documentclass[aps,prd,superscriptaddress,notitlepage,showpacs,showkeys,10pt]{revtex4-1}

\usepackage{graphicx}
\usepackage{amsmath}
\usepackage{amssymb}
\usepackage{color}
\usepackage{bm}

\definecolor{purple}{rgb}{0.8,0,0.6}
\definecolor{darkgreen}{rgb}{0.00,0.6,0.00}

\begin{document}
\title{Pseudomagnetic helicons}
\date{\today}

\author{E.~V.~Gorbar}
\affiliation{Department of Physics, Taras Shevchenko National Kiev University, Kiev, 03680, Ukraine}
\affiliation{Bogolyubov Institute for Theoretical Physics, Kiev, 03680, Ukraine}

\author{V.~A.~Miransky}
\affiliation{Department of Applied Mathematics, Western University, London, Ontario N6A 5B7, Canada}
\affiliation{Department of Physics and Astronomy, Western University, London, Ontario N6A 3K7, Canada}

\author{I. A.~Shovkovy}
\affiliation{College of Integrative Sciences and Arts, Arizona State University, Mesa, Arizona 85212, USA}
\affiliation{Department of Physics, Arizona State University, Tempe, Arizona 85287, USA}

\author{P.~O.~Sukhachov}
\affiliation{Department of Applied Mathematics, Western University, London, Ontario N6A 5B7, Canada}

\begin{abstract}
The existence of pseudomagnetic helicons is predicted for strained Dirac and Weyl materials. The corresponding
collective modes are reminiscent of the usual helicons in metals in strong magnetic fields but can exist even
without a magnetic field due to a strain-induced background pseudomagnetic field. The properties of both
pseudomagnetic and magnetic helicons are investigated in Weyl matter using the formalism of the
consistent chiral kinetic theory. It is argued that the helicon dispersion relations are affected by the electric
and chiral chemical potentials, the chiral shift, and the energy separation between the Weyl nodes.
The effects of multiple pairs of Weyl nodes are
also discussed. A simple setup for experimental detection of pseudomagnetic helicons is proposed.
\end{abstract}

\keywords{Weyl materials, chiral kinetic theory, collective excitations, helicons, pseudomagnetic field}

\pacs{71.45.-d, 03.65.Sq}

\maketitle

\section{Introduction}
\label{sec:introduction}

Electromagnetic collective excitations play an important role in various plasmas \cite{Krall,Landau:t10,Maxfield,Kaner}
including relativistic ones. The latter are usually studied in the context of the early universe \cite{Kronberg, Durrer}, the
relativistic heavy-ion collisions \cite{Kharzeev:2008-Nucl,Kharzeev:2016}, and the degenerate states of dense matter in
compact stars \cite{Kouveliotou:1999}. Since the low-energy quasiparticle excitations in the recently discovered Dirac
\cite{Borisenko,Neupane,Liu} and Weyl \cite{Tong,Bian,Qian,Long,Belopolski,Cava} materials are described by
the massless Dirac/Weyl equations, the properties of their collective effects should resemble those in relativistic plasmas.
By taking into account that the massless Dirac/Weyl quasiparticles carry a well defined chirality, an imbalance
between the number densities of opposite chirality carriers could be induced. Such a chiral asymmetry opens
the possibility of qualitatively new effects and could modify the properties of collective excitations in the relativistic-like
quasiparticle plasma.

It is worth pointing out that Dirac and Weyl materials may not only reveal some characteristic properties
of truly relativistic forms of matter but also allow one to probe absolutely new quantum effects inaccessible in high-energy
physics. Some of them, for example, are connected with the unusual plasma response to a background pseudomagnetic
field $\mathbf{B}_5$. Unlike an ordinary magnetic field $\mathbf{B}$, a pseudomagnetic one couples with
different signs to fermions of different chiralities. In Weyl and Dirac materials, the pseudomagnetic field can be induced by
various types of strains \cite{Zubkov:2015,Cortijo:2016yph,Cortijo:2016,Grushin-Vishwanath:2016,Pikulin:2016,Liu-Pikulin:2016}.
In the case of Cd$_3$As$_2$ material, for example, it is estimated that the magnitude of the corresponding field could reach
about $B_5\approx0.3~\mbox{T}$ when a static torsion is applied to a nanowire \cite{Pikulin:2016} and about
$B_5\approx15~\mbox{T}$ when a thin film is bent \cite{Liu-Pikulin:2016}. Since Weyl nodes in condensed matter
materials always come in pairs of opposite chirality (this stems from the Nielsen--Ninomiya theorem \cite{Nielsen-Ninomiya}),
the pseudomagnetic field by itself does not break the time-reversal symmetry in Weyl materials.

Plasmons are perhaps the best known and characteristic collective excitations in a plasma. They are gapped excitations whose
minimal energy is determined by the Langmuir (plasma) frequency. Recently, we showed \cite{Gorbar:2016ygi,Gorbar:2016sey} that plasmons in
relativistic matter in constant magnetic and pseudomagnetic fields are, in fact, {\em chiral} (pseudo)magnetic plasmons. Their
chiral nature is manifested in the oscillations
of the chiral charge density, which are absent for ordinary electromagnetic plasmons. Moreover, the constant
pseudomagnetic field $\mathbf{B}_{5}$  affects the dependence of plasmon frequencies already in the linear
order in the wave-vector. Similar modifications to the energy dispersion of these plasmons can be also induced by the
chiral shift parameter $\mathbf{b}$ (i.e., the momentum-space separation of the Weyl nodes) in Weyl materials.
Interestingly, even in the absence of the chiral shift and external fields, the chiral chemical potential leads to a
splitting of the plasmon frequencies in the linear order in the wave vector \cite{Gorbar:2016sey}.

\begin{figure}[t]
\begin{center}
\includegraphics[width=0.8\textwidth]{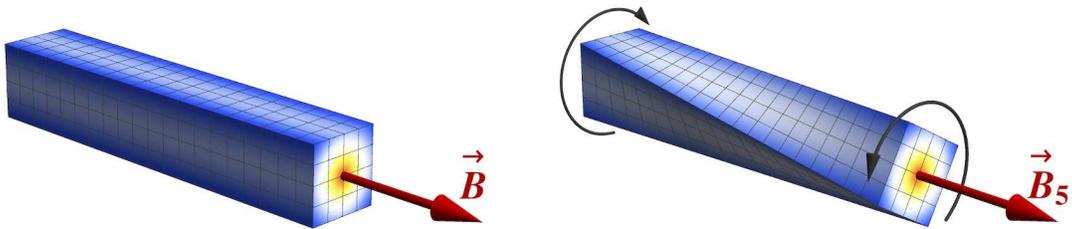}
\caption{The illustration of two limiting setups for observing helicons. While the usual helicons can be induced in
Weyl or Dirac materials in a background magnetic field (left panel), the pseudomagnetic helicons should exist in Weyl or Dirac
materials with torsion/strain-induced pseudomagnetic field (right panel).}
\label{fig:illustration}
\end{center}
\end{figure}

For a long time it was believed that low-frequency electromagnetic waves cannot propagate in metals. However,
the authors of Refs.~\cite{Konstantinov:1960,Aigrain:1960} showed that there exist transverse low-energy
gapless excitations propagating along the background magnetic field in uncompensated metals (i.e., metals
with different electron and hole densities), which were called helicons. Their counterparts propagating in
ionospheres of planets are known as whistlers. The pioneer study of helicons in Weyl materials was
performed in Ref.~\cite{Pellegrino}, where it was shown that the dispersion law of these collective
excitations encodes information on the chiral shift parameter $\mathbf{b}$. In this paper, we extend
the corresponding study to the case of a background pseudomagnetic field, a nonzero chiral chemical
potential, and temperature. The two limiting setups for observing helicons with magnetic $\mathbf{B}$ and pseudomagnetic $\mathbf{B}_{5}$ fields are schematically illustrated in the left and right panels in Fig.~\ref{fig:illustration}, respectively. Using the consistent chiral kinetic theory, we show that {\em pseudomagnetic helicon} can exists in Weyl and Dirac materials under strain.

The paper is organized as follows. The polarization vector and the characteristic equation for the
low-energy collective excitations are derived in Sec.~\ref{sec:collective-B}. The pseudomagnetic
helicons are analyzed in Sec.~\ref{sec:helicon-result}. In Sec.~\ref{sec:multipair} we
generalize the study of pseudomagnetic helicons to the case of Weyl materials with multiple pairs
of Weyl nodes, as well as to the case of Dirac materials. The main results are summarized in
Sec.~\ref{sec:Summary-Discussions}. In Appendix \ref{sec:CKT}, we present the key
details of the consistent chiral kinetic theory. Some useful technical formulas and results are
given in Appendices \ref{sec:App-g} and \ref{sec:App-P}.

\section{Polarization vector and characteristic equation}
\label{sec:collective-B}

In this section, we use the consistent chiral kinetic theory \cite{Gorbar:2016ygi,Gorbar:2016sey} in
order to derive the polarization vector and the characteristic equation for the low-energy collective
excitations in Weyl materials with one pair of Weyl nodes in an effective background field $\mathbf{B}_{0,\lambda}\equiv
\mathbf{B}_0+\lambda \mathbf{B}_{0,5}$, where $\mathbf{B}_0$ is an ordinary magnetic field,
$\mathbf{B}_{0,5}$ is a strain-induced pseudomagnetic field, and $\lambda=\pm$ is the
fermion chirality. It should be noted that the consistent formulation of the chiral kinetic theory amends
the original formulation of Refs.~\cite{Son:2012wh,Stephanov,Son-Spivak} to the case of Weyl
materials with a nonzero chiral shift parameter $\mathbf{b}$ (odd under time reversal symmetry)
and the energy separation between the Weyl nodes $b_0$ (odd under the parity transformation). Moreover, in the
presence of pseudoelectromagnetic fields, the consistent formulation of the chiral kinetic theory
\cite{Gorbar:2016ygi,Gorbar:2016sey} is required by the local conservation of the electric charge.
(For similar arguments in the context of quantum field theory, see Refs.~\cite{Landsteiner:2013sja,Landsteiner:2016}.)
Such a formulation is also essential for the correct description of the anomalous Hall
effect in Weyl materials \cite{Burkov:2011ene,Grushin-AHE,Goswami}, as well as for ensuring the absence
of the chiral magnetic effect (CME) current in an equilibrium state of Weyl matter \cite{Franz,Basar,Landsteiner:2016}.

In this study we assume that a Weyl material is subjected only to static background magnetic fields and strains.
Therefore, background pseudoelectric and electric fields are absent, $\mathbf{E}_{0}=\mathbf{E}_{0,5}=0$.
Furthermore, for simplicity, we set $\mathbf{B}_{0}\parallel \hat{\mathbf{z}}$ and $ \mathbf{B}_{0,5}\parallel\hat{\mathbf{z}}$,
where $\hat{\mathbf{z}}$ denotes the unit vector pointing in the $+z$ direction. In addition to the background
magnetic and pseudomagnetic fields, collective modes in Weyl materials may induce weak oscillating
electromagnetic fields $\mathbf{E}^{\prime}$ and $\mathbf{B}^{\prime}$. As usual in the linear regime
\cite{Krall,Landau:t10}, the corresponding fields take the form of plane waves, i.e.,
\begin{equation}
\mathbf{E}^{\prime} = \mathbf{E} e^{-i\omega t+i\mathbf{k}\cdot\mathbf{r}},
\qquad
\mathbf{B}^{\prime} = \mathbf{B} e^{-i\omega t+i\mathbf{k}\cdot\mathbf{r}},
\end{equation}
with the frequency $\omega$ and the wave vector $\mathbf{k}$. The Maxwell equations imply
that $\mathbf{B}^{\prime} = c(\mathbf{k}\times \mathbf{E}^{\prime})/\omega$ and
\begin{eqnarray}
\mathbf{k}\left(\mathbf{k}\cdot \mathbf{E}^{\prime} \right) - k^2 \mathbf{E}^{\prime}
=-\frac{\omega^2}{c^2}\left(n_0^2\mathbf{E}^{\prime}+4\pi \mathbf{P}^{\prime}\right),
\label{collective-B-tensor-spectrum-EM-0}
\end{eqnarray}
where
\begin{equation}
P^{\prime m}\equiv P^{m} e^{-i\omega t+i\mathbf{k}\cdot\mathbf{r}}= i\frac{J^{\prime m}}{\omega}=\chi^{ml}E^{\prime l}
\label{collective-B-polarization-def}
\end{equation}
is the polarization vector, $J^{\prime m}$ is the oscillating part of the current, see Eq.~(\ref{consistent-4-current}), $\chi^{ml}$ is the electric susceptibility tensor, and $m,l=1,2,3$ denote spatial indices. Further, $n_0$ denotes the material's background refractive index, which in the case of Dirac semimetal Cd$_3$As$_2$ is $n_0\approx6$ \cite{Freyland}.
In order to simplify our analysis here, we will neglect the dependence of the refractive
index on the frequency. This is well justified for a relatively narrow range of frequencies relevant for helicon modes. From Eqs.~(\ref{collective-B-tensor-spectrum-EM-0}) and (\ref{collective-B-polarization-def}) we derive the following
characteristic equation:
\begin{equation}
\mbox{det}\left[\omega^2\varepsilon^{lm}- c^2k^2\delta^{lm} + c^2 k^l k^m \right]=0,
\label{collective-B-tensor-dispersion-relation-general}
\end{equation}
where
\begin{equation}
\varepsilon^{lm} = n_0^2\delta^{lm} +4\pi\chi^{lm}
\label{collective-B-tensor-dielectric}
\end{equation}
is the dielectric tensor. The dielectric tensor $\varepsilon^{lm}$ can be determined by using the consistent chiral kinetic theory.
The key details of the corresponding formalism are reviewed in Appendix~\ref{sec:CKT}. (Note that the fields $\mathbf{E}_{\lambda}$ and $\mathbf{B}_{\lambda}$ in Appendix~\ref{sec:CKT} represent the total effective electric and magnetic fields, including the background as well as oscillating ones.) In the study
of collective excitations, it is convenient to use the following anzatz for the distribution function
$f_{\lambda}=f_{\lambda}^{\rm (eq)}+f_{\lambda}^{\prime}$, where $f_{\lambda}^{\rm (eq)}$ is the equilibrium
distribution function given in Eq.~(\ref{CKT-equilibrium-function}), and
\begin{equation}
f_{\lambda}^{\prime} = f_{\lambda}^{(1)} e^{-i\omega t+i\mathbf{k}\cdot\mathbf{r}}
\end{equation}
is a perturbation related to the oscillating fields $\mathbf{E}^{\prime}$ and $\mathbf{B}^{\prime}$.
To the leading order in perturbation theory, the chiral kinetic equation (\ref{CKT-kinetic-equation}) takes the form
\begin{equation}
i\left[ (1+\kappa_{\lambda}) \omega
- (\mathbf{v}\cdot \mathbf{k})-\frac{e}{c}(\mathbf{v}\cdot\mathbf{\Omega}_{\lambda})(\mathbf{B}_{0,\lambda}\cdot \mathbf{k}) \right]
f_{\lambda}^{(1)}
- \frac{e}{c}(\mathbf{v}\times \mathbf{B}_{0,\lambda}) \cdot \partial_\mathbf{p}f_{\lambda}^{(1)}
=e\Big[ (\tilde{\mathbf{E}}\cdot\mathbf{v})
+\frac{e}{c}(\mathbf{v}\cdot\mathbf{\Omega}_{\lambda}) (\tilde{\mathbf{E}}\cdot\mathbf{B}_{0,\lambda})\Big]
\frac{\partial f_{\lambda}^{\rm (eq)}}{\partial \epsilon_{\mathbf{p}}},
\label{collective-B-kinetic-equation-13}
\end{equation}
where $\mathbf{\Omega}_{\lambda}=\lambda \hbar \mathbf{p}/(2p^3)$ is the Berry curvature \cite{Berry:1984}, $p\equiv|\mathbf{p}|$, and
$\mathbf{v}$ denotes the quasiparticle velocity, defined in Eq.~(\ref{CKT-v}) with $\mathbf{B}_{\lambda}\to \mathbf{B}_{0,\lambda}$. In the kinetic equation, we also used the
following shorthand notations:
\begin{eqnarray}
\label{collective-B-kappa-l-def}
\kappa_{\lambda} &\equiv& \frac{e}{c} (\mathbf{\Omega}_{\lambda} \cdot \mathbf{B}_{0,\lambda})
= \lambda \hbar\frac{e\left(\hat{\mathbf{p}}\cdot\mathbf{B}_{0,\lambda}\right)}{2c p^2}, \\
\label{collective-B-tilde-E-def}
\tilde{\mathbf{E}}&=& \mathbf{E}
+i\frac{\lambda v_F \hbar}{2\omega p} \mathbf{k} \left(\hat{\mathbf{p}}
\cdot[\mathbf{k}\times\mathbf{E}]\right),
\end{eqnarray}
and $\hat{\mathbf{p}}=\mathbf{p}/p$.
Note that the second term in Eq.~(\ref{collective-B-tilde-E-def}) originates from the dependence of the
quasiparticle dispersion relation on the oscillating part of the magnetic field
$\mathbf{B}^{\prime}$, i.e.,
\begin{equation}
\epsilon_{\mathbf{p}} = v_Fp\left[1 - \frac{e}{c}((\mathbf{B}_{0,\lambda}+\mathbf{B}^{\prime})\cdot \mathbf{\Omega}_{\lambda})\right].
\label{collective-B-total-epsilon}
\end{equation}
By making use of the cylindrical coordinates with the $z$-axis pointing along the effective magnetic
field $\mathbf{B}_{0,\lambda}$ and $\phi$ being the azimuthal angle of momentum $\mathbf{p}$,
Eq.~(\ref{collective-B-kinetic-equation-13}) can be rewritten in the following form:
\begin{equation}
\frac{v_FeB_{0,\lambda}}{cp}\frac{\partial f_{\lambda}^{(1)}}{\partial \phi}
+i\left[ (1-\kappa_{\lambda}) \omega
- v_F (\hat{\mathbf{p}} \cdot \mathbf{k})\right]  f_{\lambda}^{(1)} = ev_F(\hat{\mathbf{p}} \cdot \tilde{\mathbf{E}})
\frac{\partial f_{\lambda}^{\rm (eq)}}{\partial \epsilon_{\mathbf{p}}},
\label{collective-B-kinetic-equation}
\end{equation}
where we dropped all terms quadratic in $B_{0,\lambda}$. Here it is appropriate to recall that, by
construction, the chiral kinetic theory is reliable only to the linear order in $B_{0,\lambda}$ \cite{Stephanov,Son-Spivak}.

As one can see, Eq.~(\ref{collective-B-kinetic-equation}) takes the following conventional form (see, e.g., Ref.~\cite{Landau:t10}):
\begin{equation}
\frac{\partial f_{\lambda}^{(1)}}{\partial \phi} +i (a_1+a_2\cos\phi)f_{\lambda}^{(1)}
=Q(\phi),
\label{collective-B-kinetic-equation-2}
\end{equation}
where the function of the azimuthal angle on the right-hand side is given by
\begin{equation}
\label{collective-Q-be}
Q(\phi) = a_3 \cos(\phi_E-\phi) + a_4 +  a_5\frac{p_\parallel k_\parallel
+ p_\perp k_\perp \cos\phi}{p^2} \left[E_{\perp}p_{\perp}k_{\parallel}\sin{(\phi-\phi_E)} +E_{ \perp}p_{\parallel}k_{\perp}\sin{(\phi_E)}
-E_{\parallel}k_{\perp}p_{\perp}\sin{(\phi)}\right],
\end{equation}
and
\begin{equation}
a_1 = \frac{cp \omega (1-\kappa_{\lambda})}{ev_F B_{0,\lambda}} -\frac{c p_\parallel k_\parallel}{eB_{0,\lambda}}, \quad a_2 =  - \frac{c p_\perp k_\perp }{eB_{0,\lambda}}, \quad
a_3 =  \frac{c p_\perp E_{\perp}}{B_{0,\lambda}}
\frac{\partial f_{\lambda}^{\rm (eq)}}{\partial \epsilon_{\mathbf{p}}}, \quad
a_4 = \frac{c p_\parallel E_{\parallel} }{B_{0,\lambda}}
\frac{\partial f_{\lambda}^{\rm (eq)}}{\partial \epsilon_{\mathbf{p}}}, \quad
a_5 =\frac{i\lambda \hbar v_F}{2\omega B_{0,\lambda}} \frac{\partial f_{\lambda}^{\rm (eq)}}{\partial \epsilon_{\mathbf{p}}}.
\label{collective-Q-ee}
\end{equation}
Here subscripts $\parallel$ and $\perp$ denote parallel and perpendicular components of a vector with respect to the magnetic field direction
and $\phi_{E}$ denotes the azimuthal angle of $\mathbf{E}$,
which, similarly to $\phi$, is measured from the $\mathbf{k}_{\perp}$ direction in the plane perpendicular to the
magnetic field.

To the linear order in effective magnetic field strength $B_{0,\lambda}$, the equilibrium distribution function
(\ref{CKT-equilibrium-function}) can be expanded as follows:
\begin{equation}
f_{\lambda}^{\rm (eq)} \approx f^{(0)}_{\lambda}
-\frac{\lambda e v_F \hbar B_{0,\lambda}p_{\parallel}}{2 p^2 c} \frac{\partial f^{(0)}_{\lambda} }{\partial \epsilon_{\mathbf{p}} }
+ O(B_{0,\lambda}^2),
\label{collective-B-f-series}
\end{equation}
where $f^{(0)}_{\lambda}$ is the equilibrium function $f^{\rm (eq)}_{\lambda}$ at $\mathbf{B}_{0,\lambda}=0$.

As stated in the Introduction, the main goal of this study is to investigate the spectrum of (pseudo)magnetic
helicons in Weyl materials. The corresponding collective excitations are gapless modes closely related to
the cyclotron resonances. By following the same approach that is used in nonrelativistic plasmas \cite{Landau:t10},
it is convenient to replace the distribution function $f_{\lambda}^{(1)} (\phi)$ with a new function,
\begin{equation}
g(\phi) = e^{i a_2 \sin\phi} f_{\lambda}^{(1)} (\phi),
\label{helicon-f1-def}
\end{equation}
which, in view of Eq.~(\ref{collective-B-kinetic-equation-2}), satisfies the following equation:
\begin{equation}
\frac{\partial g}{\partial \phi} +i a_1g
=e^{i a_2 \sin\phi} Q(\phi).
\label{collective-B-kinetic-equation-3}
\end{equation}
By taking into account that $g(\phi)$ is a periodic function of the azimuthal angle $\phi$, the solution
to Eq.~(\ref{collective-B-kinetic-equation-3}) can be obtained in the form of a Fourier series,
\begin{equation}
g(\phi)=\sum_{n=-\infty}^{\infty} g_n e^{i n \phi},
\label{collective-B-f1-1}
\end{equation}
where coefficients $g_n$ are given by
\begin{equation}
g_n = -\frac{i}{2\pi(a_1+n)} \int_{0}^{2\pi} e^{i a_2 \sin\tau-in \tau} Q(\tau)d\tau .
\label{collective-B-gn-def}
\end{equation}
Here, the integration over the variable $\tau$ can be performed analytically. The corresponding
explicit expressions for $g_n$ are presented in Appendix~\ref{sec:App-g}.

For gapless collective excitations such as helicons, it is convenient to consider the long-wavelength limit, i.e.,
$v_Fk\ll\Omega_c|_{p=p^{*}}$, where $p^{*} \sim \sqrt{\mu_5^2 +\mu^2+\pi^2T^2}/v_F$ is a characteristic momentum
in a chiral plasma, and
\begin{eqnarray}
\Omega_c \simeq \frac{ev_F B_{0,\lambda}}{cp} +O(B_{0,\lambda}^2)
\label{collective-B-Omegac}
\end{eqnarray}
is an analog of the cyclotron frequency for massless fermions in the (pseudo)magnetic field that depends on momentum $p$.
In the long-wavelength limit, the analysis significantly simplifies because one can neglect the dependence of $f_{\lambda}^{(1)}$
on the wave vector $\mathbf{k}$. Furthermore, by utilizing the same approximation as in Ref.~\cite{Pellegrino}, we will include only
the lowest three (i.e., $n=0,\pm1$) Fourier harmonics in the solution. By using the definitions in Eq.~(\ref{collective-Q-ee}) and the
explicit expressions for the coefficients $g_n$ in Appendix~\ref{sec:App-g}, we obtain the following results:
\begin{eqnarray}
f_{\lambda, 0}^{(1)} &=& -i \frac{e v_Fp_{\parallel}(1+\kappa_{\lambda})}{p\omega} (\mathbf{E}\cdot
\hat{\mathbf{z}}) \frac{\partial f_{\lambda}^{\rm (eq)}}{\partial \epsilon_{\mathbf{p}}},
\label{collective-B-f1-k0-n0} \\
f_{\lambda, \pm}^{(1)} &=& -i\frac{ev_Fp_{\perp}(1+\kappa_{\lambda})}{2p}
\frac{E_x \mp iE_y}{\omega \pm\Omega_c} \frac{\partial f_{\lambda}^{\rm (eq)}}{\partial \epsilon_{\mathbf{p}}} e^{\pm i\phi},
\label{collective-B-f1-k0}
\end{eqnarray}
for the $n=0$ and $n=\pm1$ Fourier harmonics of $f_{\lambda}^{(1)}$, respectively. By making use of these results, we
derive the expression for the polarization vector $\mathbf{P}$, i.e.,
\begin{eqnarray}
\mathbf{P} &=&
\sum_{\lambda=\pm} \sum_{\rm p,a} \frac{ie}{\omega}\int\frac{d^3p}{(2\pi \hbar)^3} \left\{ e(\tilde{\mathbf{E}}\times\mathbf{\Omega}_{\lambda})
+\frac{e}{\omega}(\mathbf{v}\cdot\mathbf{\Omega}_{\lambda})\left( \mathbf{k} \times \mathbf{E} \right) +\frac{e}{c}(\mathbf{\delta v}
\cdot\mathbf{\Omega}_{\lambda})\mathbf{B}_{0,\lambda}\right\} f_{\lambda}^{\rm (eq)} \nonumber\\
&+&\sum_{\lambda=\pm} \sum_{\rm p,a} \frac{\lambda e^2 \hbar v_F}{2\omega^2}\int\frac{d^3p}{(2\pi \hbar)^3} \frac{1}{p} [\mathbf{k}\times
\mathbf{\Omega}_{\lambda}] \left(\hat{\mathbf{p}}\cdot[\mathbf{k}\times\mathbf{E}]\right) f_{\lambda}^{\rm (eq)}
+ \sum_{n=-1}^1\sum_{\lambda=\pm} \sum_{\rm p,a}  \frac{ie}{\omega}\int\frac{d^3p}{(2\pi \hbar)^3}\left[\mathbf{v}
+\frac{e}{c}(\mathbf{v}\cdot\mathbf{\Omega}_{\lambda})\mathbf{B}_{0,\lambda} \right] f_{\lambda, n}^{(1)} \nonumber\\
&-&\sum_{n=-1}^1\sum_{\lambda=\pm}\sum_{\rm p,a} \frac{e}{\omega} \int\frac{d^3p}{(2\pi \hbar)^3}
\epsilon_{\mathbf{p}} f^{(1)}_{\lambda, n} (\mathbf{k}\times\mathbf{\Omega}_{\lambda})
-i \frac{e^3}{2\pi^2 \omega c \hbar^2}(\mathbf{b}\times \mathbf{E}) + i \frac{e^3 b_0}{2\pi^2\omega^2 \hbar^2} (\mathbf{k}\times \mathbf{E}),
\label{collective-B-polarization}
\end{eqnarray}
where $\sum_{\rm p,a}$ denotes the summation over contributions
of particles and antiparticles (holes), and we used the expression for the current consistent with the local charge conservation
\cite{Gorbar:2016ygi,Gorbar:2016sey}. Here,
\begin{equation}
\delta\mathbf{v} = \frac{2ev_F}{c} \hat{\mathbf{p}}\left(\mathbf{B}\cdot\mathbf{\Omega}_{\lambda}\right)
-\frac{ev_F}{c}\mathbf{B}\left(\hat{\mathbf{p}}\cdot\mathbf{\Omega}_{\lambda}\right)
=\frac{\lambda \hbar ev_F}{2\omega p^2} \Big\{2\hat{\mathbf{p}}\left(\hat{\mathbf{p}}\cdot[\mathbf{k}\times\mathbf{E}]\right)
-[\mathbf{k}\times\mathbf{E}]\Big\}
\label{collective-B-delta-v}
\end{equation}
is the correction to velocity,
which follows from the oscillating magnetic field in the dispersion relation (\ref{collective-B-total-epsilon}).
The details of calculation of the polarization vector $\mathbf{P}$ in the limit of small frequencies $\omega \ll \Omega_{c}|_{p=p^{*}}$
are given in Appendix \ref{sec:App-P}. The final result in the leading order in $B_{0,\lambda}$ takes the following form:
\begin{eqnarray}
4\pi\mathbf{P} = A_1 (\mathbf{E}\times\hat{\mathbf{z}}) +A_2(\hat{\mathbf{k}}\times \mathbf{E}) +A_3(\mathbf{b}\times \mathbf{E}) +A_4 (\mathbf{E}-\hat{\mathbf{z}}(\mathbf{E}\cdot\hat{\mathbf{z}})) +A_5\hat{\mathbf{z}}(\mathbf{E}\cdot\hat{\mathbf{z}}),
\label{collective-B-P-total}
\end{eqnarray}
where
\begin{eqnarray}
\label{collective-B-Gs-be}
A_1&=& \sum_{\lambda=\pm} i\frac{2e c \mu_{\lambda}}{3B_{0,\lambda}v_F^3\hbar^3\pi \omega} \left(\mu_{\lambda}^2+\pi^2T^2\right) \equiv i\frac{\tilde{A}_1}{\omega},\\
A_2&=&i \frac{2k e^2 (eb_0+\mu_5)}{\pi\omega^2 \hbar^2} \equiv i \frac{k \tilde{A}_2}{\omega^2},\\
\label{collective-B-Gs-A3}
A_3&=& -i \frac{2e^3}{\pi \omega c \hbar^2} \equiv i\frac{\tilde{A}_3}{\omega},\\
A_4&=& \sum_{\lambda=\pm} \frac{2 c^2}{3\pi\hbar^3 v_F^5B_{0,\lambda}^2}
\left(\mu_{\lambda}^4 +2\pi^2 \mu_{\lambda}^2T^2+\frac{7\pi^4 T^4}{15}\right) ,\\
A_5&=& -\sum_{\lambda=\pm}\frac{2e^2}{3\pi\hbar^3v_F \omega^2}
\left(\mu_{\lambda}^2 +\frac{\pi^2T^2}{3}\right)\equiv\frac{\tilde{A}_5}{\omega^2}.
\label{collective-B-Gs-ee}
\end{eqnarray}
At zero temperature, we obtain
\begin{eqnarray}
\tilde{A}_1&\stackrel{T\to0}{=}&
\sum_{\lambda=\pm}\frac{2e c \mu_{\lambda}^3}{3\pi\hbar^3 B_{0,\lambda} v_F^3},\\
A_4&\stackrel{T\to0}{=}&\sum_{\lambda=\pm}  \frac{2c^2\mu_{\lambda}^4}{3\pi\hbar^3B_{0,\lambda}^2v_F^5} ,\\
\tilde{A}_5&\stackrel{T\to0}{=}&-\sum_{\lambda=\pm}  \frac{2e^2 \mu_{\lambda}^2}{3\pi \hbar^3 v_F}.
\label{collective-B-Gs-T=0}
\end{eqnarray}
(Note that the other two coefficients, i.e., $\tilde{A}_2$ and $\tilde{A}_3$, do not depend on temperature.)
Thus, the dielectric tensor (\ref{collective-B-tensor-dielectric}) reads
\begin{eqnarray}
\varepsilon^{ml}=\delta^{ml}n_0^2 +A_1 \epsilon^{ml3}
+A_2 \epsilon^{mjl} \hat{\mathbf{k}}^j +A_3\epsilon^{mjl} \mathbf{b}^j  +A_4 (\delta^{ml}-\delta^{m3}\delta^{l3})
+A_5\delta^{m3}\delta^{l3}.
\label{collective-B-chi}
\end{eqnarray}

It is worth noting that contrary to the case of the usual helicons in metals, the dielectric tensor in Weyl materials is
modified by the chiral shift $\mathbf{b}$, as can be seen from the fourth term in Eq.~(\ref{collective-B-chi}). The dielectric tensor is also
affected by the pseudomagnetic field $\mathbf{B}_{0,5}$ and the chiral chemical potential $\mu_5$. By taking
into account that $b_0=-\mu_5/e$ in equilibrium, we find that $\varepsilon^{ml}$ is symmetric with respect to the
replacement $\left(\mathbf{B}_{0,5}, \mu_5\right) \rightarrow \left(\mathbf{B}_0, \mu\right)$. It is also worth
noting that the interband (i.e., particle-hole) contributions, which may be important in the optical range, can be
ignored in the study of low-energy helicons. In fact, the corresponding effects can be effectively accounted
for by renormalizing the background refractive index $n_0$ \cite{Pellegrino}.

It is well known that helicons are absent at $\mathbf{k}\perp\mathbf{B}_0$ \cite{Landau:t10, Krall, Maxfield,Kaner}.
Therefore, for simplicity, we can set $\mathbf{k}=\left(0,0,k_{\parallel}\right)$.
Then, the characteristic equation (\ref{collective-B-tensor-dispersion-relation-general}) takes the form
\begin{equation}
A_3^2 \omega^4 b_{\perp}^2\left(c^2 k^2-(n_0^2+A_4) \omega^2\right) -(n_0^2+A_5)\left[\left(c^2 k^2-(n_0^2+A_4) \omega^2\right)^2
+ \omega^4 \left(A_1-A_2-b_{\parallel}A_3\right)^2\right] =0.
\label{collective-B-dispersion-relation-long}
\end{equation}
Before solving this equation, let us briefly discuss the role of coefficients $A_{i}$
(where $i=\overline{1,5}$). As in the case of usual helicons in metals, the existence of pseudomagnetic ones
relies on the off-diagonal components of the dielectric tensor, which are given by coefficients $A_1$, $A_2$, and,
in the case of $\mathbf{b} \ne 0$, by $A_3$. The first coefficient is related to the nondissipative
Hall conductivity, albeit generalized to the case of nonzero values of the pseudomagnetic field and
the chiral chemical potential.
The coefficient $A_2$ is proportional to the combination of the energy separation of the Weyl nodes and the chiral
chemical potential, $eb_0+\mu_5$, that vanishes in equilibrium \cite{Landsteiner:2013sja,Landsteiner:2016}.
While in our analysis below we will eventually assume the state of equilibrium, it is interesting to note that the
helicon properties could be substantially modified in Weyl materials out of equilibrium, e.g., in steady states with
$\mathbf{E}_0\cdot \mathbf{B}_0\neq 0$, which are characterized by $eb_0+\mu_5\neq 0$. The third term, $A_3$,
is related to the anomalous Hall effect in Weyl materials \cite{Burkov:2011ene,Grushin-AHE,Goswami}.
The other two coefficients, i.e., $A_4$ and $A_5$, affect only diagonal components of the dielectric tensor
$\varepsilon^{ml}$ and, thus, are not crucial for the existence of helicons. However, they could provide quantitative
corrections to the dispersion relations of collective excitations.

\section{Low-energy collective modes}
\label{sec:helicon-result}

In this section we study the helicon-type solutions to the characteristic equation (\ref{collective-B-dispersion-relation-long})
and investigate their properties. The analysis of the corresponding equation shows that due to the large factor $(n^2_0+A_5)$,
the effect of the chiral shift perpendicular to a background (pseudo)magnetic field $\mathbf{b}_{\perp}$ is
numerically small for the helicon dispersion relation. Therefore,
we consider below only the case $\mathbf{b}=\left(0,0,b_{\parallel}\right)$, which admits simple analytical solutions for the
collective excitation frequencies. Then, Eq.~(\ref{collective-B-dispersion-relation-long})
reduces to the following equation:
\begin{equation}
\left(n_0^2\omega^2-c^2k^2+\omega^2A_4\right)^2 +\omega^4 (A_1-A_2-A_3b_{\parallel})^2=0,
\label{helicon-dispersion-relation-long-bz}
\end{equation}
where we also dropped the overall factor $(n_0^2+A_5)$ because the equation $(n_0^2+A_5)=0$
has only a high-energy gapped solution with frequency proportional to the Langmuir (plasma) frequency
\begin{equation}
\Omega_e \equiv \sqrt{\frac{4\alpha}{3\pi\hbar^2}\left(\mu^2+\mu_5^2 +\frac{\pi^2 T^2}{3}\right)}.
\label{helicon-Langmuir-def}
\end{equation}
Here $\alpha\equiv e^2/(\hbar v_F)$ is the fine structure constant. In view of our approximation of small $\omega$, this
solution is unreliable. The corresponding gapped excitations (namely, the chiral magnetic plasmons) were properly analyzed
in Refs.~\cite{Gorbar:2016ygi,Gorbar:2016sey}.

The solutions to Eq.~(\ref{helicon-dispersion-relation-long-bz}) are
\begin{eqnarray}
\omega_{\pm}=\frac{\pm |\tilde{A}_1-\tilde{A}_3b_{\parallel}|+\sqrt{4k \left(c^2k\mp\tilde{A}_2\right)(n_0^2+A_4)
+\left(\tilde{A}_3b_{\parallel}- \tilde{A}_1\right)^2}}{2(n_0^2+A_4)},
\label{helicon-Tnot0-omega-All-sum-s-b0}
\end{eqnarray}
where coefficients $\tilde{A}_1$, $\tilde{A}_2$, $\tilde{A}_3$, and $A_4$ are given by Eqs.~(\ref{collective-B-Gs-be})
through (\ref{collective-B-Gs-ee}). In the long-wavelength limit, we find
\begin{eqnarray}
\label{helicon-Tnot0-omega-All-app-sum-s-b0-1}
\omega_{+}&\simeq& \frac{\left|\tilde{A}_1-\tilde{A}_3b_{\parallel}\right|}{n^2_0+A_4}
- \frac{\tilde{A}_2 k}{\left|\tilde{A}_1-\tilde{A}_3b_{\parallel}\right|} +k^2\frac{c^2 \left|\tilde{A}_1-\tilde{A}_3b_{\parallel}
\right|^2-\tilde{A}_2^2(n_0^2+A_4)}{\left|\tilde{A}_1-\tilde{A}_3b_{\parallel}\right|^3}+O(k^3), \\
\label{helicon-Tnot0-omega-All-app-sum-s-b0-2}
\omega_{-}&\simeq& \frac{\tilde{A}_2 k}{\left|\tilde{A}_1-\tilde{A}_3b_{\parallel}\right|} +k^2\frac{c^2\left| \tilde{A}_1-\tilde{A}_3b_{\parallel}
\right|^2-\tilde{A}_2^2(n_0^2+A_4)}{\left|\tilde{A}_1-\tilde{A}_3b_{\parallel}\right|^3}+O(k^3).
\end{eqnarray}
The gapped solution $\omega_{+}$ is unreliable since it is outside the validity of the low-frequency approximation
used in the derivation. The frequency of the gapless mode is physical and corresponds to a helicon, i.e.,
$\omega_h=\omega_{-}$. In the limit of zero temperature, $T\to0$, the corresponding result reads
\begin{eqnarray}
\omega_{h} &\simeq&\frac{3k (eb_0+\mu_5)ce\hbar v_F^3(B_0^2-B_{0,5}^2)}
{2B_0c^2\mu(\mu^2+3\mu_5^2) -2B_{0,5}c^2\mu_5(\mu_5^2+3\mu^2)+3(B_0^2-B_{0,5}^2)e^2\hbar v_F^3b_{\parallel}}
\nonumber\\
&+& \frac{3\pi k^2 c^3 \hbar^3 v_F^3 (B_0^2 -B_{0,5}^2)  }
{2e\left[2B_0c^2\mu(\mu^2+3\mu_5^2) -2B_{0,5}c^2\mu_5(\mu_5^2+3\mu^2) +3(B_0^2-B_{0,5}^2)e^2\hbar v_F^3b_{\parallel}\right]}
\nonumber\\
&-& \frac{9k^2 (eb_0+\mu_5)^2c e \hbar^2 v_F^4 (B_0^2-B_{0,5}^2)\left[B_0^2(\mu^4+6\mu^2\mu_5^2+\mu_{5}^4) +B_{0,5}^2(\mu^4+6\mu^2\mu_5^2+\mu_{5}^4) -8B_0B_{0,5}\mu\mu_5(\mu^2+\mu_5^2)\right]}{2\left[B_0\mu(\mu^2+3\mu_5^2) -B_{0,5}\mu_5(\mu_5^2+3\mu^2)\right]^2 \left[2B_0c^2\mu(\mu^2+3\mu_5^2) -2B_{0,5}c^2\mu_5(\mu_5^2+3\mu^2)+9(B_0^2+B_{0,5}^2)e^2 \hbar v_F^3 b_{\parallel}\right]}
+O(k^3),\nonumber\\
\label{helicon-omega-app-sum-s-0}
\end{eqnarray}
where we kept only the leading and subleading terms in $B_0$ and $B_{0,5}$ in the numerators and denominators.
It is instructive to consider two special cases
\begin{eqnarray}
\label{helicon-omega-app-sum-s-1}
\omega_{h}\Big|_{B_{0,5}\to0, \mu_5\to0} &\simeq& \frac{2b_0 B_{0} c e^4 v_F^2k}
{\pi c^2\hbar^2 \mu \Omega_e^2+2B_0e^4v_F^2b_{\parallel}} + \frac{eB_0c^3\hbar^2\pi v_F^2 k^2}
{\pi \hbar^2 c^2 \Omega_e^2\mu +2B_0 e^4 v_F^2 b_{\parallel}} -\frac{4B_0 c e^7 v_F^2k^2b_0^2}
{\pi \hbar^2 \Omega_e^2(\pi c^2\hbar^2 \mu \Omega_e^2 +6B_0e^4v_F^2b_{\parallel})}\nonumber\\
&\stackrel{b_0\to0}{=}& \frac{eB_0c^3\hbar^2\pi v_F^2 k^2}{\pi \hbar^2 c^2 \Omega_e^2\mu +2B_0 e^4 v_F^2 b_{\parallel}} +O(k^3),\\
\label{helicon-omega-app-sum-s-2}
\omega_{h}\Big|_{B_{0}\to0, \mu\to0} &\simeq& \frac{2(eb_0+\mu_5) B_{0,5} c e^3 v_F^2k}
{\pi c^2\hbar^2 \mu_5 \Omega_e^2+2B_{0,5}e^4v_F^2b_{\parallel}}
+ \frac{eB_{0,5}c^3\hbar^2\pi v_F^2 k^2}{\pi \hbar^2 c^2 \Omega_e^2\mu_5 +2B_{0,5} e^4 v_F^2 b_{\parallel}}
-\frac{4B_{0,5} c e^5 v_F^2k^2(eb_0+\mu_5)^2}{\pi \hbar^2 \Omega_e^2(\pi c^2\hbar^2 \mu_5 \Omega_e^2 +6B_{0,5}e^4v_F^2b_{\parallel})}\nonumber\\
&\stackrel{b_0\to-\mu_5/e}{=}& \frac{eB_{0,5}c^3\hbar^2\pi v_F^2 k^2}{\pi \hbar^2 c^2 \Omega_e^2\mu_5 +2B_{0,5} e^4 v_F^2 b_{\parallel}} +O(k^3).
\end{eqnarray}
Note that Eq.~(\ref{helicon-omega-app-sum-s-1}) agrees with the results obtained in Ref.~\cite{Pellegrino}. It should
be emphasized, though, that the default choice of parameters in Ref.~\cite{Pellegrino}, i.e., $b_0\neq 0$ and $\mu_5=0$,
effectively describes an out-of-equilibrium state of a Weyl plasma. In such a regime, the dispersion is linear in the wave vector.
On the other hand, we find that the assumption of equilibrium generically implies a quadratic dispersion relation for the helicon,
i.e., it is qualitatively the same as in the usual metals. This is due to the fact that the linear term in
Eq.~(\ref{helicon-omega-app-sum-s-0}) is proportional to $eb_0+\mu_5$ and, thus, vanishes in equilibrium. In essence,
this is the same argument that explains the absence of the CME current in Weyl materials in equilibrium
\cite{Franz,Basar,Landsteiner:2016}.

One of the key predictions of this paper is the existence of a gapless helicon-type mode in a Weyl matter without a
magnetic field. It is natural to call the corresponding gapless
mode a pseudomagnetic helicon. Indeed, as we see from Eq.~(\ref{helicon-omega-app-sum-s-2}), a gapless mode can be naturally
realized in parity-odd Weyl materials under strain. The corresponding materials are characterized
by a strain-induced background pseudomagnetic field $\mathbf{B}_{0,5}$ and a nonzero chiral
chemical potential $\mu_5$. In equilibrium, the latter is determined by the energy separation
between the Weyl nodes, i.e., $\mu_5=-eb_0$.
In such a state of Weyl materials, the helicon has a quadratic dispersion relation, see the second line in
Eq.~(\ref{helicon-omega-app-sum-s-2}). On the other hand, out of equilibrium (e.g., in a steady state with $eb_0+\mu_5\neq 0$ produced
by external fields with $\mathbf{E}_0\cdot \mathbf{B}_0\neq 0$), a linear dispersion relation
could be realized too.

In order to discuss the qualitative properties of the pseudomagnetic helicons, it is convenient to define a characteristic
scale for the chiral shift in Weyl materials. To this end, let us introduce the following reference value:
\begin{equation}
eb^{*}=0.3 \frac{\pi \hbar v_F}{c_3},
\end{equation}
where $c_3\approx 25.480~\mbox{\AA}$ is the lattice spacing and $b^{*}$ is comparable to the momentum space separation
between the Dirac points in Cd$_3$As$_2$ \cite{Neupane}. In what follows, we will concentrate only on the equilibrium case $\mu_5=-eb_0$ with nonzero
pseudomagnetic field $\mathbf{B}_{0,5}$ and set $\mathbf{B}_{0}=0$. Also, in our numerical calculations below, we
will use the value of the Fermi velocity of Cd$_3$As$_2$ \cite{Neupane}, i.e., $v_F\approx1.5 \times10^{8}~\mbox{cm/s}$.
It is expected, of course, that the main qualitative conclusions should remain valid for generic Dirac or Weyl materials.

\begin{figure}[!ht]
\begin{center}
\includegraphics[width=0.45\textwidth]{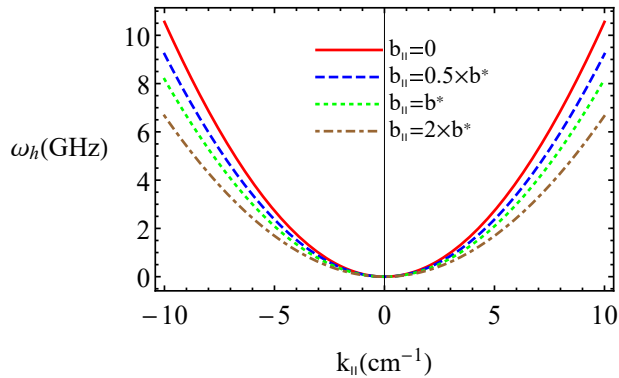}
\caption{The helicon dispersion relation $\omega_{h}=\omega_{-}$ given by Eq.~(\ref{helicon-Tnot0-omega-All-sum-s-b0})
for $b_{\parallel}=0$ (red solid line), $b_{\parallel}=0.5\,b^{*}$ (blue dashed line), $b_{\parallel}=b^{*}$ (green dotted line),
and $b_{\parallel}=2\,b^{*}$ (brown dot-dashed line). We set $B_{0,5}=10^{-2}~\mbox{T}$, $B_0=0$, $\mu_5=5~\mbox{meV}$,
and $\mu=0$.}
\label{fig:helicon-long-bz-app-sum-s}
\end{center}
\end{figure}

The effect of the chiral shift parameter $b_{\parallel}$ on the gapless mode $\omega_{h}$ is shown in
Fig.~\ref{fig:helicon-long-bz-app-sum-s}. As one can see, the chiral shift decreases the helicon
frequency. Quantitatively, however, the effect is rather weak.

\begin{figure}[!ht]
\begin{center}
\includegraphics[width=0.45\textwidth]{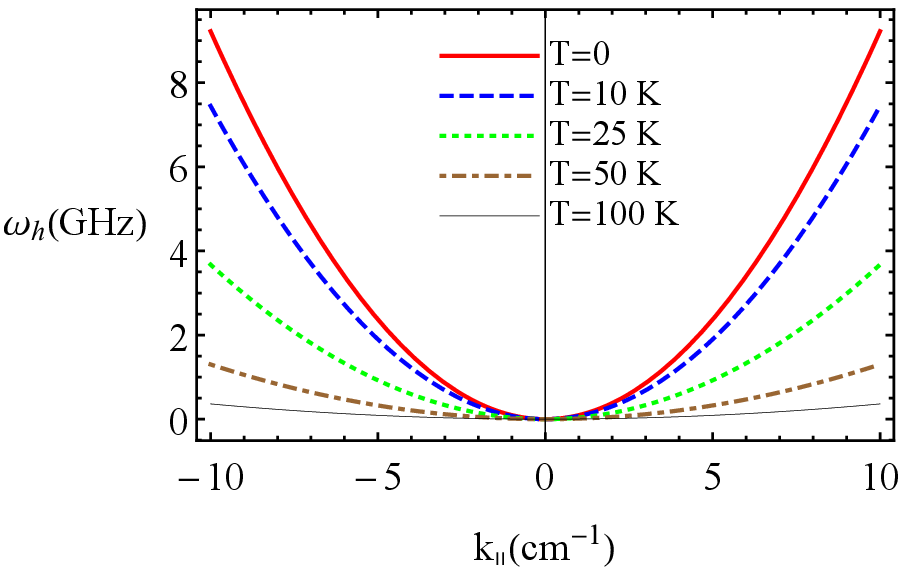} \hfill
\includegraphics[width=0.45\textwidth]{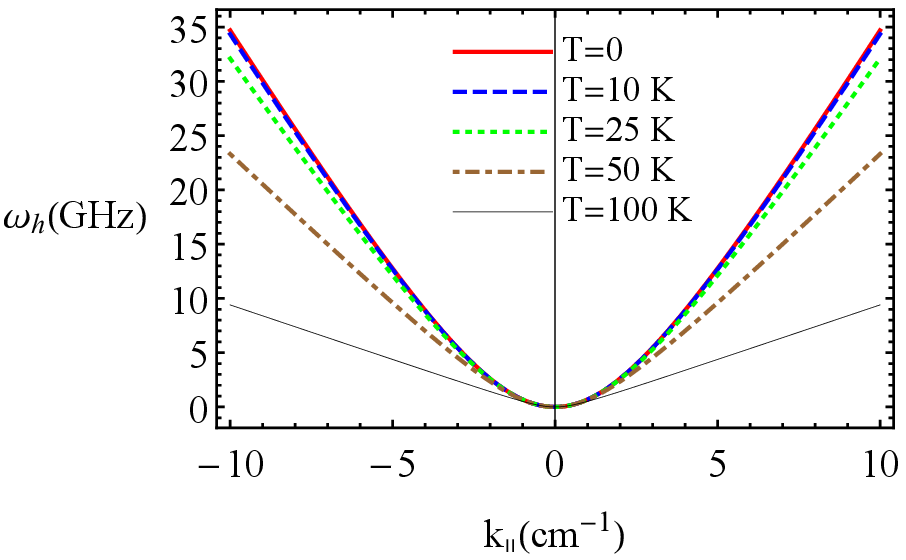}
\caption{The frequency of the collective mode $\omega_{h}=\omega_{-}$ defined by Eq.~(\ref{helicon-Tnot0-omega-All-sum-s-b0}).
Red solid, blue dashed, green dotted, brown dot-dashed, and black thin solid lines correspond to $T=0$, $T=10~\mbox{K}$,
$T=25~\mbox{K}$, $T=50~\mbox{K}$, and $T=100~\mbox{K}$, respectively. The left panel is plotted for $\mu_5=5~\mbox{meV}$
and $\mu=0$, while the right one represents results obtained for $\mu=5~\mbox{meV}$ and $\mu_5=0$.
We set $b_{\parallel}=0.5\,b^{*}$, $B_0=0$, and $B_{0,5}=10^{-2}~\mbox{T}$.}
\label{fig:helicon-long-bz-k-Tnot0-sum-s}
\end{center}
\end{figure}

\begin{figure}[!ht]
\begin{center}
\includegraphics[width=0.45\textwidth]{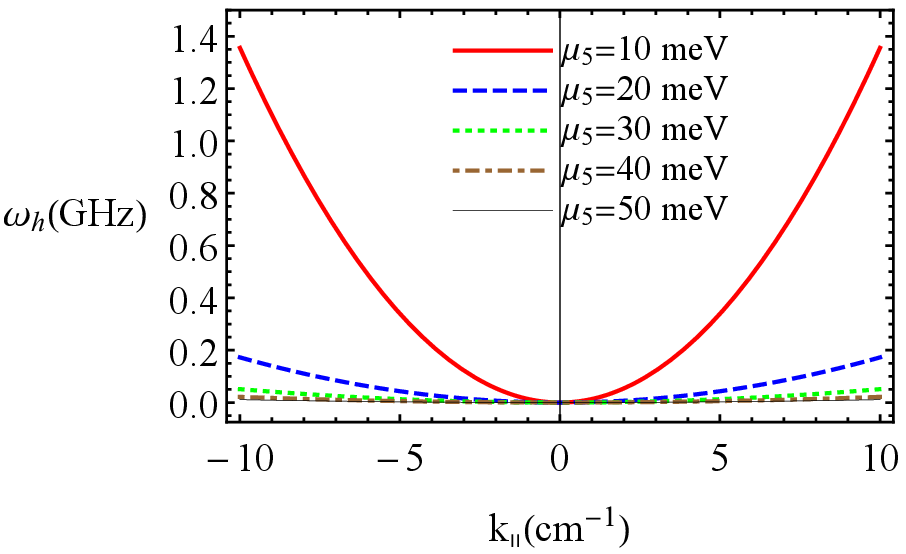} \hfill
\includegraphics[width=0.45\textwidth]{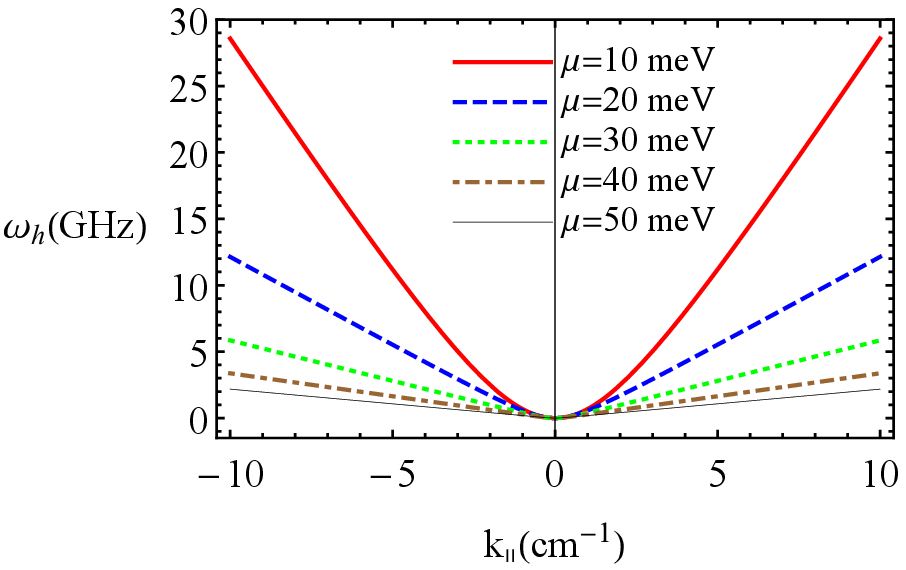}
\caption{The dispersion relation of the low-energy collective mode $\omega_{h}=\omega_{-}$ defined by Eq.~(\ref{helicon-Tnot0-omega-All-sum-s-b0}). Red solid, blue dashed, green dotted, brown dot-dashed, and black thin solid lines in the left panel correspond to $\mu_5=10~\mbox{meV}$, $\mu_5=20~\mbox{meV}$, $\mu_5=30~\mbox{meV}$, $\mu_5=40~\mbox{meV}$, and $\mu_5=50~\mbox{meV}$, respectively. The same values, albeit for the electric chemical potential $\mu$, are used in the right panel. The left panel is plotted for $\mu=0$, while the right one represents results obtained for $\mu_5=0$. We set $b_{\parallel}=0.5\,b^{*}$, $T=0$, $B_0=0$, and $B_{0,5}=10^{-2}~\mbox{T}$.}
\label{fig:helicon-long-bz-mu-mu5-Tnot0-sum-s}
\end{center}
\end{figure}

\begin{figure}[!ht]
\begin{center}
\includegraphics[width=0.45\textwidth]{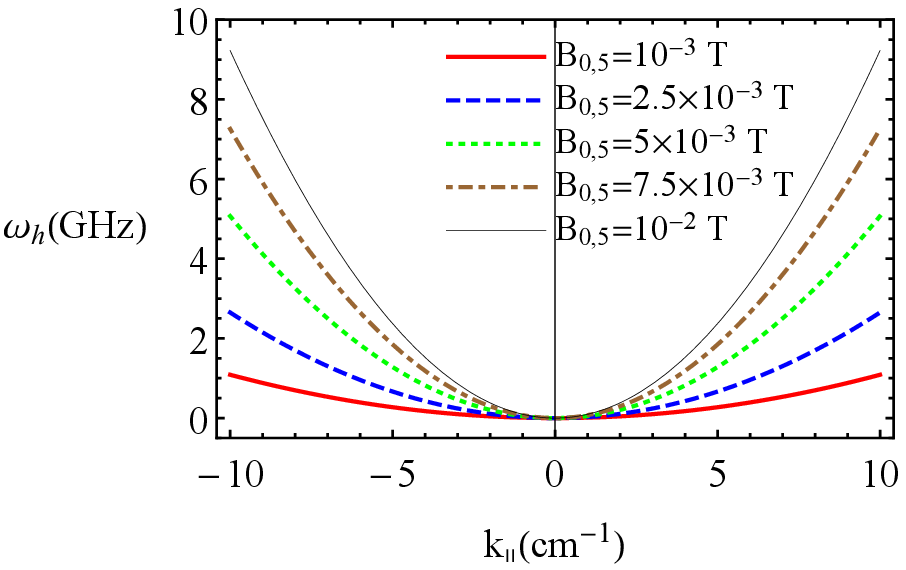} \hfill
\includegraphics[width=0.45\textwidth]{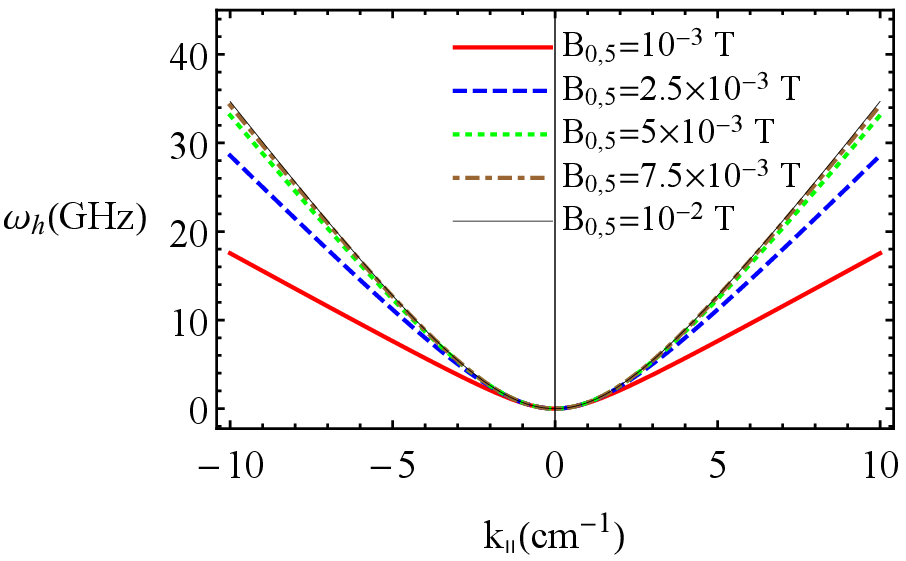}
\caption{The dispersion relation of the low-energy collective mode $\omega_{h}=\omega_{-}$ defined by
Eq.~(\ref{helicon-Tnot0-omega-All-sum-s-b0}). Red solid, blue dashed, green dotted, brown dot-dashed,
and black thin solid lines correspond to $B_{0,5}=10^{-3}~\mbox{T}$, $B_{0,5}=2.5 \times10^{-3}~\mbox{T}$,
$B_{0,5}=5\times10^{-3}~\mbox{T}$, $B_{0,5}=7.5\times10^{-3}~\mbox{T}$, and $B_{0,5}=10^{-2}~\mbox{T}$,
respectively. The left panel is plotted for $\mu_{5}=5~\mbox{meV}$, $\mu=0$, while the right one represents
results obtained for $\mu_5=0$, $\mu=5~\mbox{meV}$. We set $b_{\parallel}=0.5\,b^{*}$, $T=0$, and $B_0=0$.}
\label{fig:helicon-long-bz-B5-Tnot0-sum-s}
\end{center}
\end{figure}

The helicon dispersion relations at different values of $T$, $B_{0,5}$, $\mu$, and $\mu_5$ are plotted in
Figs.~\ref{fig:helicon-long-bz-k-Tnot0-sum-s} through \ref{fig:helicon-long-bz-B5-Tnot0-sum-s}.
Comparing the left and the right panels in Fig.~\ref{fig:helicon-long-bz-k-Tnot0-sum-s} one can see that
the dispersion law of the pseudomagnetic helicons changes  from the quadratic form at $\mu_{5}\neq0$, $\mu=0$, and small $k$ to a linear one at $\mu_{5}=0$, $\mu\neq0$, and large $k$. In the latter case, the approximate, quadratic
in $k$, expression (\ref{helicon-Tnot0-omega-All-app-sum-s-b0-2}) is valid only for small $k$. Moreover, the frequency of the
helicon mode in the left panel is a few times lower than in the right one.

Further, we show the dependence of the frequency $\omega_{h}$ on the wave vector $k$ at different
values of chiral and electric chemical potentials in the left and right panels in Fig.~\ref{fig:helicon-long-bz-mu-mu5-Tnot0-sum-s},
respectively. In order to be consistent with the approximation of small pseudomagnetic fields
\begin{equation}
\frac{\hbar v_F^2|eB_{0,5}|}{c(\mu_{5}^2+\mu^2+\pi^2T^2)}\ll1,
\end{equation}
we consider only relatively small pseudomagnetic fields $B_{0,5}\lesssim \bar{B}_{5}$ or sufficiently large electric and chiral
chemical potentials $\mu, \mu_5\gtrsim \bar{\mu}$, where
\begin{eqnarray}
\bar{B}_{5}&=&\frac{c(\mu_{5}^2+\mu^2+\pi^2T^2)}{e\hbar v_F^2}\stackrel{T\to0, \mu\to0}{=} \frac{c\mu_5^2}{e\hbar v_F^2}\approx 6.853\times10^{-4} (\mu_5[\mbox{meV}])^2~\mbox{T}, \\
\bar{\mu} &=& v_F\sqrt{\frac{\hbar |eB_{0,5}|}{c}} \approx 38.198\sqrt{B_{0,5}[\mbox{T}]}~\mbox{meV}.
\end{eqnarray}
Similarly to the right panel in Fig.~\ref{fig:helicon-long-bz-k-Tnot0-sum-s}, the dispersion law of the pseudomagnetic
helicons at $\mu_{5}=0$ and $\mu\neq0$ (right panel) changes from a quadratic one at small $k$ to a linear one at large $k$.
The dependence of the frequency $\omega_{h}$ on the wave vector $k$ at different values of the pseudomagnetic
field is shown in the left and right panels in Fig.~\ref{fig:helicon-long-bz-B5-Tnot0-sum-s} for $\mu_{5}=5~\mbox{meV}$,
$\mu=0$ and $\mu_5=0$, $\mu=5~\mbox{meV}$, respectively. Compared to the quadratic decrease of the helicon frequency
with $\mu_5$ and $\mu$, the dependence of $\omega_{h}$ on $B_{0,5}$ is almost linear.
Finally, we would like to note that the results for a finite background magnetic field $\mathbf{B}_0$ applied to the system with
nonzero chemical potential $\mu$ can be obtained by replacing
$\left(\mathbf{B}_{0,5}, \mu_5\right) \rightarrow \left(\mathbf{B}_0, \mu\right)$.

Up to now we studied helicons in a Weyl material with a single pair of Weyl nodes. However, all experimentally
discovered Weyl materials \cite{Tong,Bian,Qian,Long,Belopolski,Cava} have multiple pairs of Weyl nodes.
It is natural, therefore, to investigate how the physical properties of helicons are affected by the presence
of several pairs of Weyl nodes. Another important question is the existence of the pseudomagnetic helicons
in Dirac materials. Naively, by taking into account the topological triviality of a Dirac point, one may simply suggest
that there should be no pseudomagnetic field and, thus, no pseudomagnetic helicons in Dirac materials. However,
some Dirac semimetals (e.g., A$_3$Bi where A$=$Na,K,Rb) are, in fact, $\mathbb{Z}_2$ Weyl semimetals
\cite{Gorbar:2014sja}, whose Dirac points come from two superimposed pairs of Weyl nodes with opposite chiral
shifts. In such a case, a strain-induced pseudomagnetic field is possible and, as a result, pseudomagnetic helicons
can exist. The properties of pseudomagnetic helicons in the case of Weyl and Dirac materials with multiple pairs of
Weyl nodes are studied in the next section.

\section{Effects of multiple pairs of Weyl nodes}
\label{sec:multipair}

In order to study how the presence of several pairs of Weyl nodes affects the physical properties of
pseudomagnetic helicons, in this section we consider simple models of Weyl materials with
two and three pairs of Weyl nodes, as well as a model of Dirac material with two Dirac points.
Formally, the presence of several pairs of Weyl nodes could be taken into account by adding the pair
index $\xi$ to the one-particle distribution functions, i.e., $f_{\lambda} \to f_{\lambda}^{(\xi)}$. Note that, in
general, the pairs of Weyl nodes are characterized by different values of the chiral shift $\mathbf{b}^{(\xi)}$
and the energy separation $b_0^{(\xi)}$. By taking this into account and adding the partial contributions due
to each pair of Weyl nodes, it is straightforward to derive the expression
for the polarization vector. The final result will have the same form as in Eq.~(\ref{collective-B-polarization}),
but will include an additional sum over the pair index $\xi$.

While the case of a purely magnetic field could be straightforwardly analyzed by using the results in the previous
sections without qualitative changes, the case of a strained Weyl material is different. Indeed, contrary
to the usual magnetic field, the pseudomagnetic one is connected with the chiral shift
\cite{Cortijo:2016yph,Pikulin:2016, Liu-Pikulin:2016} and consequently depends on the pair index.
In view of different orientations of $\mathbf{b}^{(\xi)}$, in general, quasiparticles of the corresponding
pairs experience pseudomagnetic fields of different directions and amplitudes.
This makes the analysis of strained multipair Weyl or Dirac materials somewhat more complicated.

Let us begin our study with the general expression for the axial vector potential of an arbitrary pair of Weyl nodes in a strained
material, i.e.,
\begin{equation}
\label{collective-M-A5-0}
\left(\mathbf{A}^5_i\right)^{(\xi)}= - \frac{c}{v_F}\left\{\sum_{j=1}^3\left(u_{ij}-\delta_{ij}u_{jj}\right)b_j^{(\xi)}
+ \frac{\hbar^2 v_F^2 b_{i}^{(\xi)} u_{ii}}{e^2c_i^2 \left(b^{(\xi)}\right)^2}\right\}
\end{equation}
where $\mathbf{u}$ is a displacement vector, $u_{ij}\equiv(\partial_iu_j+\partial_ju_i)/2$ is a strain tensor, and $c_i$ is a lattice constant
in $i$ direction. The corresponding pseudomagnetic field equals
\begin{equation}
\label{collective-M-B5-0}
\left(\mathbf{B}_{0,5}^{(\xi)}\right)_i= - \frac{c}{v_F}\left\{ \frac{1}{2}\sum_{j,k,l=1}^3\varepsilon_{ijk}\left(\partial_j\partial_lu_{k}-2\delta_{kl}\partial_j\partial_lu_{l}
\right)b_l^{(\xi)} + \sum_{j,k=1}^3\varepsilon_{ijk}\frac{\hbar^2 v_F^2 \partial_j\partial_ku_{k}b_{k}^{(\xi)}}{e^2c_k^2 \left(b^{(\xi)}\right)^2}\right\}.
\end{equation}
The amplitude and the direction of this field strongly depend on the directions of the chiral shifts, as well as
on the type of strain. Therefore, in the following subsections we will consider three simplified cases: (i) static
torsion along the $+z$ axis in a Weyl material, (ii) static bending along the $+y$ axis of a Weyl material, and
(iii) static torsion along the $+z$ axis in a Dirac material. For the sake of simplicity, we assume that an external
magnetic field is absent.

\subsection{Weyl materials with torsion}
\label{sec:multipair-torsion}

The static torsion along the $+z$ axis is characterized by the following displacement vector:
\begin{equation}
\label{collective-M-u-t}
\mathbf{u}=u_0 z [\mathbf{r}\times\hat{\mathbf{z}}],
\end{equation}
where $u_0$ is a constant which depends on the magnitude of a torsion and sample details. Let us begin with the simplest model of a Weyl material with two pairs of Weyl nodes. The first pair is
characterized by the chiral shift
$\mathbf{b}^{(1)}=\left(0, 0, b_{z}\right)$. Further, we assume that the chiral shift
$\mathbf{b}^{(2)}=\left(b_{x}, 0, 0\right)$ of the second pair is orthogonal to $\mathbf{b}^{(1)}$ because the case of parallel
$\mathbf{b}^{(1)}$ and $\mathbf{b}^{(2)}$ is rather trivial.

\begin{figure}[t]
\begin{center}
\includegraphics[width=0.45\textwidth]{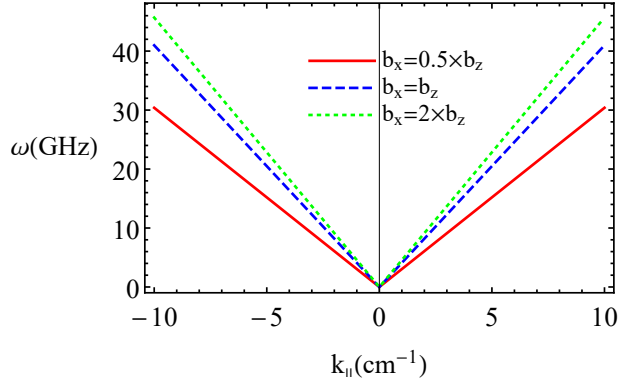}
\caption{The dependence of the gapless collective mode frequency on the wave vector obtained by solving
Eq.~(\ref{collective-B-tensor-dispersion-relation-general}) with dielectric tensor (\ref{collective-M-chi-t}) for $b_{x}=0.5\,b_{z}$
(red solid line), $b_{x}=b_{z}$ (blue dashed line), and $b_{x}=2\,b_{z}$ (green dotted line). We set
$B_{0,5}=10^{-2}~\mbox{T}$, $B_0=0$, $eb_0^{(1)}=eb_0^{(2)}=-\mu_5^{(1)}=-\mu_5^{(2)}=-5~\mbox{meV}$, $b_{z}=b^{*}$, and $\mu=0$.}
\label{fig:helicon-M}
\end{center}
\end{figure}

Using Eq.~(\ref{collective-M-B5-0}), one can easily find the following pseudomagnetic fields for the
quasiparticles in the vicinity of the first and second pairs of Weyl nodes:
\begin{eqnarray}
\label{collective-M-B5-t-be}
\mathbf{B}_{0,5}^{(1)}&=&  \frac{cu_0 b_{z}}{v_F} \hat{\mathbf{z}}
\equiv B_{0,5}\hat{\mathbf{z}}, \\
\mathbf{B}_{0,5}^{(2)}&=& -\frac{B_{0,5} b_x}{2b_{z}} \hat{\mathbf{x}},
\label{collective-M-B5-t-ee}
\end{eqnarray}
respectively.
It is worth noting that the quasiparticles of the second pair of Weyl nodes experience the pseudomagnetic field directed opposite to the chiral shift $\mathbf{b}^{(2)}$. Their contribution to the polarization vector reads
\begin{eqnarray}
4\pi\mathbf{P}^{(2)} = A_1^{(2)} (\mathbf{E}\times\hat{\mathbf{x}}) +A_2^{(2)}(\hat{\mathbf{k}}\times \mathbf{E}) +A^{(2)}_3(\mathbf{b}^{(2)}
\times \mathbf{E}) +A_4^{(2)} (\mathbf{E}-\hat{\mathbf{x}}(\mathbf{E}\cdot\hat{\mathbf{x}}))
+A_5^{(2)}\hat{\mathbf{x}}(\mathbf{E}\cdot\hat{\mathbf{x}}).
\label{collective-M-P-t}
\end{eqnarray}
Here coefficients $A_j^{(\xi)}$ with $j=\overline{1,5}$ and $\xi=1,2$ are given by Eqs.~(\ref{collective-B-Gs-be}) through
(\ref{collective-B-Gs-ee})
with $\mu\to\mu_{\lambda}^{(\xi)}=\mu+\lambda\mu_5^{(\xi)}$, $b_0\to b_0^{(\xi)}$, and $B_{0,\lambda} \to B_{0,\lambda}^{(\xi)} =\lambda B_{0,5}^{(\xi)}$.
Then, we easily find the total dielectric tensor in the model under consideration:
\begin{eqnarray}
\varepsilon^{ml}&=&\delta^{ml}n_0^2+4\pi\left(\chi^{(1)}\right)^{ml}+4\pi\left(\chi^{(2)}\right)^{ml}=\delta^{ml}n_0^2 +A_1^{(1)} \epsilon^{ml3}
+A_1^{(2)} \epsilon^{ml1}
+\left(A_2^{(1)}+A_2^{(2)}\right) \epsilon^{mjl} \hat{\mathbf{k}}^j  \nonumber\\
&+&A_3\left(b_{z}\epsilon^{m3l}+b_{x}\epsilon^{m1l}\right) +A_4^{(1)} \left(\delta^{ml}-\delta^{m3}\delta^{l3}\right) +A_4^{(2)}
\left(\delta^{ml}-\delta^{m1}\delta^{l1}\right)
+A_5^{(1)}\delta^{m3}\delta^{l3} +A_5^{(2)}\delta^{m1}\delta^{l1}.
\label{collective-M-chi-t}
\end{eqnarray}
By solving the characteristic equation (\ref{collective-B-tensor-dispersion-relation-general}) with
the dielectric tensor (\ref{collective-M-chi-t}), we find that there are gapless collective excitations in the
strained Weyl material with two pairs of Weyl nodes whose frequency dependence on the wave vector for different
values of $b_{x}$ is plotted in Fig.~\ref{fig:helicon-M}. As one can see, the frequency of the gapless
mode in the long-wavelength limit depends linearly on the wave vector even in equilibrium, i.e.,
at $\mu_5^{(1)}=-eb_0^{(1)}$ and $\mu_5^{(2)}=-eb_0^{(2)}$.
The corresponding dependence can be approximated as
\begin{equation}
\label{collective-M-omega-t}
\omega \simeq v_1 |k| +O(k^3),
\end{equation}
where
\begin{eqnarray}
\label{collective-M-omega-1-t}
v_1 &=& \tilde{A}_5^{(1)}\left[c^2\tilde{A}_5^{(2)} +\left(\tilde{A}_2^{(1)}+\tilde{A}_2^{(2)}\right)^2\right] \Bigg\{\tilde{A}_5^{(1)}
\left(\tilde{A}_2^{(1)}+\tilde{A}_2^{(2)}\right)\left(\tilde{A}_1^{(1)}-b_z\tilde{A}_3\right) +\sqrt{\tilde{A}_5^{(1)}} \Bigg[\tilde{A}_5^{(1)}
\left(\tilde{A}_2^{(1)}+\tilde{A}_2^{(2)}\right)^2\left(\tilde{A}_1^{(1)}-b_z\tilde{A}_3\right)^2 \nonumber\\ &+&\left(c^2\tilde{A}_5^{(2)}+
\left(\tilde{A}_2^{(1)}+\tilde{A}_2^{(2)}\right)^2\right) \left(\tilde{A}_5^{(1)}\tilde{A}_5^{(2)}\left(n_0^2+\tilde{A}_4^{(1)}
+\tilde{A}_4^{(2)}\right) -\tilde{A}_5^{(1)}\left(\tilde{A}_1^{(1)}-b_z\tilde{A}_3\right)^2 -\tilde{A}_5^{(2)}\left(\tilde{A}_1^{(2)}-b_x
\tilde{A}_3\right)^2\right)\Bigg]^{1/2}\Bigg\}^{-1}.\nonumber\\
\end{eqnarray}
This result shows that the presence of the second pair of Weyl nodes does affect the physical properties of
pseudomagnetic helicons in a nontrivial way, namely, the quadratic dependence of the frequency on the
wave vector is replaced by a linear one.

The effect becomes even more dramatic when another pair of Weyl nodes (with a chiral shift perpendicular to
the other two) is added to the model. To see this, let us assume that there are three pairs of Weyl nodes with
the chiral shifts orthogonal to each other, i.e., $\mathbf{b}^{(1)}=\left(0, 0, b_{z}\right)$,
$\mathbf{b}^{(2)}=\left(b_x, 0, 0\right)$, and $\mathbf{b}^{(3)}=\left(0, b_y, 0\right)$. Under the same torsion,
the pseudomagnetic field (\ref{collective-M-B5-0}) for the quasiparticles in the vicinity of the third pair
of Weyl nodes is
\begin{equation}
\label{collective-M-B5-3}
\mathbf{B}_{0,5}^{(3)}= -\frac{B_{0,5} b_{y}}{2 b_{z}}\hat{\mathbf{y}}.
\end{equation}
The corresponding contribution to the polarization vector reads
\begin{eqnarray}
4\pi\mathbf{P}^{(3)} = A_1^{(3)} (\mathbf{E}\times\hat{\mathbf{y}}) +A_2^{(3)}(\hat{\mathbf{k}}\times \mathbf{E}) +A_3(\mathbf{b}^{(3)}
\times \mathbf{E}) +A_4^{(3)} (\mathbf{E}-\hat{\mathbf{y}}(\mathbf{E}\cdot\hat{\mathbf{y}}))
+A_5^{(3)}\hat{\mathbf{y}}(\mathbf{E}\cdot\hat{\mathbf{y}}).
\label{collective-M-P-total-3}
\end{eqnarray}
By taking into account the contributions from all three pairs of Weyl nodes in the characteristic equation
(\ref{collective-B-tensor-dispersion-relation-general}), we find that the corresponding collective mode
becomes gapped. In other words, there is no gapless helicon in the spectrum.
The case of a Weyl material with more than three pairs of nodes can be straightforwardly analyzed by using the
results obtained here.

\subsection{Weyl materials with bending}
\label{sec:multipair-banding}

Let us briefly discuss the case where a pseudomagnetic field is induced by bending the film of a Weyl
material with three pairs of mutually orthogonal Weyl nodes: $\mathbf{b}^{(1)}=\left(0, 0, b_{z}\right)$,
$\mathbf{b}^{(2)}=\left(b_{x}, 0, 0\right)$, and $\mathbf{b}^{(3)}=\left(0, b_{y}, 0\right)$. By using
the model of Ref.~\cite{Liu-Pikulin:2016}, we assume that the bending along the $+y$ axis can be
described by the following displacement vector:
\begin{equation}
\label{collective-M-u-band}
\mathbf{u}=u_1 xz\hat{\mathbf{z}},
\end{equation}
where $u_1$ is a constant which depends on the magnitude of banding and sample properties.
From the general result in Eq.~(\ref{collective-M-B5-0}), we find the following pseudomagnetic fields for
all three Weyl nodes:
\begin{eqnarray}
\label{collective-M-B5-band-be}
\mathbf{B}_{0,5}^{(1)}&=& cu_1\left(\frac{v_F\hbar^2}{c_3^2e^2b_{z}} -\frac{b_z}{2v_F}\right)\hat{\mathbf{y}}
\equiv \tilde{B}_{0,5}\hat{\mathbf{y}},\\
\mathbf{B}_{0,5}^{(2)}&=& \mathbf{B}_{0,5}^{(3)}=0.
\label{collective-M-B5-band-ee}
\end{eqnarray}
Note that a nonzero pseudomagnetic field is produced only for the pair with $\mathbf{b}\parallel
\hat{\mathbf{z}}$ and its direction is orthogonal to the chiral shift. This is qualitatively different from the case
of torsion discussed in the previous subsection, where the pseudomagnetic fields were either along or against
the directions of the chiral shifts, see Eqs.~(\ref{collective-M-B5-t-be}), (\ref{collective-M-B5-t-ee}), and
(\ref{collective-M-B5-3}). In the case of bending, on the other hand, the two pairs with the chiral shifts orthogonal to the
$+z$ axis do not experience any pseudomagnetic field at all.

Thus, in the model at hand, the polarization vector for the first pair is given by Eq.~(\ref{collective-M-P-total-3})
with the replacement $A_j^{(3)}\to A_j^{(1)}$ ($j=\overline{1,5}$), $\mathbf{b}^{(3)}\to\mathbf{b}^{(1)}$,
$\mu_{\lambda}^{(3)}\to\mu_{\lambda}^{(1)}=\mu+\lambda\mu_{5}^{(1)}$, and $B_{0,\lambda}^{(3)}\to B_{0,\lambda}^{(1)}
=\lambda \tilde{B}_{0,5}$. By making use of the results from our previous study \cite{Gorbar:2016sey},
we find that coefficients (\ref{collective-B-Gs-be})
through (\ref{collective-B-Gs-ee}) for the second and third pairs in the long-wavelength limit equal
\begin{eqnarray}
\label{collective-M-B-Gs-be-band}
A_1^{(\xi)}&=& 0,\\
A_2^{(\xi)}&=&i \frac{2k e^2 (3eb_0^{(\xi)}+2\mu_5^{(\xi)})}{3\pi\omega^2 \hbar^2} \equiv i \frac{k \tilde{A}_2^{(\xi)}}{\omega^2},\\
A_3^{(\xi)}&=& A_3^{(1)}=A_3,\\
A_4^{(\xi)}&=& -\frac{4e^2}{3\pi\hbar^3v_F \omega^2}\left(\mu^2+\left(\mu_5^{(\xi)}\right)^2 +\frac{\pi^2 T^2}{3}\right)
\equiv \frac{\tilde{A}_4^{(\xi)}}{\omega^2},\\
A_5^{(\xi)}&=& A_4^{(\xi)}, \quad \tilde{A}_5^{(\xi)}= \tilde{A}_4^{(\xi)},
\label{collective-M-B-Gs-ee-band}
\end{eqnarray}
where $\xi=2,3$. Note that while $A_1^{(\xi)}=0$, the coefficients $A_4^{(\xi)}$ for the
second and third pairs depend on frequency. This has a profound effect on the properties
of collective modes, i.e., they become gapped. There are three solutions to
Eq.~(\ref{collective-B-tensor-dispersion-relation-general}). One of them describes a longitudinal
mode with the gap of the order of the Langmuir frequency. The other two are
\begin{eqnarray}
\omega_{\pm} &\simeq& \frac{\sqrt{(A_{1,1}-A_{3,1}b_{\parallel})^2 -4A_{4,2}(n_0^2+A_{4,0})} \mp(A_{1,1}-A_{3,1}b_{\parallel})}{2(n_0^2+A_{4,0})}
\pm k \frac{A_{2,2}}{\left[(A_{1,1}-A_{3,1}b_{\parallel})^2 -4A_{4,2}(n_0^2+A_{4,0})\right]^{1/2}} \nonumber\\ &+&k^2\frac{c^2\left[(A_{1,1}-
A_{3,1}b_{\parallel})^2 -4A_{4,2}(n_0^2+A_{4,0})\right]-A_{2,2}^2(n_0^2+A_{4,0})}{\left[(A_{1,1}-A_{3,1}b_{\parallel})^2 -4A_{4,2}
(n_0^2+A_{4,0})\right]^{3/2}} +O(k^3).
\label{helicon-M-Tnot0-omega-band-ee}
\end{eqnarray}
Here, for simplicity, we ignored the small corrections proportional to $b_x$ and $b_y$, and introduced
the following notation:
\begin{eqnarray}
\label{collective-M-B-Aij-be-band}
A_{1,1} &\equiv& \sum_{\xi=1}^3 \tilde{A}_1^{(\xi)},\\
A_{2,2} &\equiv& \sum_{\xi=1}^3 \tilde{A}_2^{(\xi)},\\
A_{3,1} &\equiv& 3\tilde{A}_3,\\
A_{4,0} &\equiv& A_4^{(1)},\\
A_{4,2} &\equiv& \tilde{A}_4^{(2)}+\tilde{A}_4^{(3)},\\
A_{5,2} &\equiv& \sum_{\xi=1}^3 \tilde{A}_5^{(\xi)}.
\label{collective-M-B-Aij-ee-band}
\end{eqnarray}
By analyzing Eq.~(\ref{helicon-M-Tnot0-omega-band-ee}), we find that the transverse modes are also gapped and
their energies are reminiscent of the transverse plasma frequencies $\omega^{\pm}_{\rm tr}$ of the chiral
pseudomagnetic plasmons \cite{Gorbar:2016sey}.

In conclusion, the physical properties of pseudomagnetic helicons in a mutipair Weyl material are qualitatively
different from those with a single pair of Weyl nodes. They are unlike the magnetic helicons, for which the multipair
case is a trivial generalization of a single pair one.

\subsection{Dirac materials}
\label{sec:multipair-Dirac}

Let us consider the pseudomagnetic helicons in the experimentally relevant model of the Dirac material with two Dirac points, e.g., for A$_3$Bi (where A$=$Na,K,Rb). In view of the $\mathbb{Z}_2$ structure of such a model \cite{Gorbar:2014sja}, it can be viewed as a superposition of two pairs of Weyl nodes with opposite chiral shifts: $\mathbf{b}^{(1)}=\left(0, 0, b_{z}\right)$ and $\mathbf{b}^{(2)}=\left(0, 0, -b_z\right)$. Then, as follows from the analysis in Secs.~\ref{sec:multipair-torsion} and \ref{sec:multipair-banding}, strains produce equal in magnitude but opposite in direction pseudomagnetic fields for each pair of the Weyl nodes.
The polarization vector for the first pair is given by Eq.~(\ref{collective-B-P-total}) with replacement $A_j\to A_j^{(1)}$ ($j=\overline{1,5}$), where $\mu\to\mu_{\lambda}^{(1)}=\mu+\lambda\mu_5^{(1)}$, $b_0\to b_0^{(1)}$, and $B_{0,\lambda} \to \lambda B_{0,5}$. The polarization vector for the other pair is given by a similar expression, but with the replacement $A_j\to A_j^{(2)}$ ($j=\overline{1,5}$), $\mu\to\mu_{\lambda}^{(2)}=\mu+\lambda\mu_5^{(2)}$, $b_0\to b_0^{(2)}$, and $B_{0,\lambda} \to -\lambda B_{0,5}$.
Then, the total dielectric tensor in the Dirac material is
\begin{eqnarray}
\varepsilon^{ml}=\delta^{ml}n_0^2 +\left(A_1^{(1)}+A_1^{(2)}\right) \epsilon^{ml3}
+\left(A_2^{(1)}+A_2^{(2)}\right) \epsilon^{mjl} \hat{\mathbf{k}}^j  +\left(A_4^{(1)}+A_4^{(2)}\right)\left(\delta^{ml}-\delta^{m3}\delta^{l3}\right)
+\left(A_5^{(1)}+A_5^{(2)}\right)\delta^{m3}\delta^{l3}, \nonumber\\
\label{collective-M-chi-Dirac}
\end{eqnarray}
where the $+z$ axis can be viewed as the direction of the pseudomagnetic field.

Let us begin with a general case, where Dirac points are separated in energy similarly to the Weyl ones. In this case $b_0^{(1)}=-b_0^{(2)}=b_0$ and $\mu_5^{(1)}=-\mu_5^{(2)}=\mu_5$. Then, using Eqs.~(\ref{collective-B-Gs-be}) through (\ref{collective-B-Gs-ee}), we find that $A_j^{(1)}= A_j^{(2)}=A_j$ ($j=1, 4, 5$) and $A_2^{(1)}= -A_2^{(2)}=A_2$. This situation corresponds to the equilibrium one considered in Sec.~\ref{sec:helicon-result}, albeit with trivial replacements $A_j\to 2A_j$ ($j=1, 4, 5$) and $A_2\to 0$. Then, using Eq.~(\ref{helicon-dispersion-relation-long-bz}) at $b_{\parallel}=0$ we find the following frequencies of collective modes:
\begin{eqnarray}
\label{collective-M-omega-pm-Dirac-ii}
\omega_{\pm} &=& \frac{\sqrt{\tilde{A}_1^2+c^2k^2\left(n_0^2+2A_4^2\right)}\pm|\tilde{A}_1|}{n_0^2+2A_4},
\end{eqnarray}
As is easy to check, the frequency $\omega_{+}$ describes a gapped high-energy mode,
which cannot be described reliably  in the current low-energy formalism. The frequency of the other
mode in the long-wavelength limit takes the following form:
\begin{eqnarray}
\label{collective-M-omega-pm-Dirac-2-ii}
\omega_{-}&\simeq& \frac{c^2k^2 B_{0,5} 3v_F^3\hbar^3\pi}{\left|\sum_{\lambda=\pm} 4e c \lambda\mu_{\lambda} \left(\mu_{\lambda}^2+\pi^2T^2\right)\right|} +O(k^3).
\end{eqnarray}
This defines a pseudomagnetic helicon with quadratic dispersion relation. It is similar to the pseudomagnetic
helicon in the Weyl material with a one pair of Weyl nodes [cf. with Eq.~(\ref{helicon-Tnot0-omega-All-app-sum-s-b0-2})].

In the absence of the energy separation between the Dirac points, i.e., at $b_0^{(1)}=b_0^{(2)}=0$ and $\mu_5^{(1)}=\mu_5^{(2)}=0$, coefficient $\tilde{A}_1=0$ and both frequencies  in the long-wavelength limit read
\begin{eqnarray}
\omega_{+}=\omega_{-}=\frac{ck}{\sqrt{n_0^2+2A_4}} \simeq \frac{ck B_{0,5} \sqrt{3\pi\hbar^3 v_F^5}}{\sqrt{3\pi\hbar^3 v_F^5 n_0^2B_{0,5}^2+8c^2
\left(\mu^4 +2\pi^2 \mu^2T^2+7\pi^4 T^4/15\right) }}.
\label{collective-M-omega-pm-Dirac-bz-eq-i}
\end{eqnarray}
As in a Weyl material with two pairs of Weyl nodes, we have a linear dispersion relation of the pseudomagnetic helicons.

Before concluding this section, it is important to reiterate that pseudomagnetic helicons can exist in most
Dirac materials under strain. That is due to the fact that such compounds \cite{Borisenko,Neupane,Liu} usually have
a single pair of Dirac points, which are equivalent to two pairs of Weyl nodes with opposite chiral shifts. Moreover,
compared to experimentally discovered Weyl materials, which typically have a large number of Weyl nodes (e.g., TaAs has 12
such pairs \cite{Tong,Bian,Qian,Long}), Dirac materials may in fact be a much more convenient platform for studying the pseudomagnetic helicons.

\section{Summary and discussions}
\label{sec:Summary-Discussions}

By making use of the consistent chiral kinetic theory, we obtained and analyzed the spectrum of the low-energy
gapless helicon-type modes in strained Weyl materials. Unlike the usual helicons, these collective excitations
exist even in the absence of a background magnetic field. The necessary ingredients for the existence of these
helicons are a strain-induced pseudomagnetic field $\mathbf{B}_{0,5}$ and a chiral chemical potential $\mu_5$.
Note that the latter appears naturally in the equilibrium state of a parity-odd Weyl material with a nonzero energy
separation $b_0$ between the Weyl nodes. We call this type of collective excitation a \emph{pseudomagnetic
helicon}.

We found that in the equilibrium state with $\mu_5=-eb_0$, when both $\mathbf{B}_{0,5}$ and $\mu_5$ are present,
the pseudomagnetic helicon has a conventional, quadratic dependence of its frequency on the wave vector $\mathbf{k}$.
The situation changes qualitatively in the case $B_{0,5}\neq0$ ($B_{0}\neq0$) but $\mu_5=0$ ($\mu=0$), where
the dispersion becomes approximately linear in $k$. We suggest also that linear dispersion relations for helicons
are possible in the out-of-equilibrium states of Weyl materials with $eb_0+\mu_5\neq 0$. The
corresponding (steady) states of matter could be induced, for example, by applying external electromagnetic fields with
$\mathbf{E}_0\cdot \mathbf{B}_0\neq 0$. In this paper, we also studied the effects of temperature, as well as electric and
chiral chemical potentials on the properties of helicons. We found, in particular, that all three of them have a tendency to
decrease the helicon frequency for a given wave vector.

It is worth noting that the necessary ingredients for the existence of pseudomagnetic helicons are naturally
present in Weyl materials, making them ideal platforms to study the anomalous physics. Indeed, the
pseudomagnetic field $\mathbf{B}_{0,5}$ can be induced by a strain and the chiral chemical potential
$\mu_5$ in equilibrium is determined by the energy separation between Weyl nodes. The effect of the
chiral shift $\mathbf{b}$ on the pseudomagnetic helicon is weak, but may be detectable via the change
of its frequency.

Further, we showed that the gapless collective modes can exist also in Weyl materials with many pairs of Weyl nodes, as well as in some Dirac materials. For the
simplest model
of a multipair Weyl material with two pairs of Weyl nodes under the static torsion, we found that the frequency of these modes is
linear in the wave vector even in equilibrium. This result is clearly different from the usual, quadratic in $k$, dispersion law of the
magnetic helicons in Weyl materials with many pairs of nodes or their pseudomagnetic counterparts for a single pair of Weyl nodes.
Moreover, for a Weyl material with three pairs of Weyl nodes and mutually orthogonal chiral shifts, gapless modes are absent. The
pseudomagnetic helicons are also absent when strains do not produce pseudomagnetic field for a certain pair of Weyl nodes. Moreover, we investigated the case of Dirac materials with two Dirac points. Owing to the nontrivial $\mathbb{Z}_2$ structure, the pseudomagnetic helicons can propagate in these materials. The dispersion law of such helicons ranges from quadratic in $k$ when the Dirac points are separated in energy to linear when such a separation is absent. The case of a general Weyl matter with many pairs of Weyl nodes in the presence of pseudomagnetic as well as background magnetic fields deserves a further in-depth investigation. While such an analysis could be done, in principle, by using
the results of the present study, it is rather cumbersome and will be reported elsewhere.

Last but not least, we would like to propose a simple experimental setup that should allow one
to detect pseudomagnetic helicons. Based on the same idea that is used in metals, it requires
measuring the amplitude of transmission of an electromagnetic wave through a Weyl or Dirac crystal as a function
of an applied strain (which can be quantified by a bending or torsion angle), or as a function of the frequency at a fixed strain. Because of an
interference of standing helicon waves inside the sample, the resulting signal should oscillate
as the function of strain, after the strain reaches a sufficiently large magnitude. Note that overcoming
a critical value of the strain corresponds to entering the regime of a sufficiently large pseudocyclotron
frequency compared to the value of the pseudomagnetic helicon frequency. Indeed, this is the condition for the existence of
pseudomagnetic helicons that can propagate without suffering too much damping. In such a setup, it
is also possible to study the effects of the chiral shift parameter by changing the orientation of the
crystal and/or by applying strains along different directions.

\begin{acknowledgments}
The work of E.V.G. was partially supported by the Program of Fundamental Research of the Physics and
Astronomy Division of the NAS of Ukraine.
The work of V.A.M. and P.O.S. was supported by the Natural Sciences and Engineering Research Council of Canada.
The work of I.A.S. was supported by the U.S. National Science Foundation under Grant No.~PHY-1404232.
\end{acknowledgments}

\appendix

\section{Equations of the consistent chiral kinetic theory}
\label{sec:CKT}

In this appendix, we briefly review the main aspects of the consistent chiral kinetic theory considered in
Refs.~\cite{Gorbar:2016ygi,Gorbar:2016sey}. The time evolution of one-particle distribution functions
$f_{\lambda}(t,\mathbf{p},\mathbf{r})$ for the fermions of chirality $\lambda=\pm$ are governed in the chiral
kinetic theory \cite{Stephanov,Son-Spivak} by the following equation in the collisionless limit:
\begin{equation}
\partial_tf_{\lambda}+\frac{1}{1+\frac{e}{c}(\mathbf{B}_{\lambda}\cdot\mathbf{\Omega}_{\lambda})}
\left[\Big(e\tilde{\mathbf{E}}_{\lambda}
+\frac{e}{c}(\mathbf{v}\times \mathbf{B}_{\lambda})
+\frac{e^2}{c}(\tilde{\mathbf{E}}_{\lambda}\cdot\mathbf{B}_{\lambda})\mathbf{\Omega}_{\lambda}\Big)\cdot
\partial_\mathbf{p}f_{\lambda}
+\Big(\mathbf{v}+e(\tilde{\mathbf{E}}_{\lambda}\times\mathbf{\Omega}_{\lambda})
+\frac{e}{c}(\mathbf{v}\cdot\mathbf{\Omega}_{\lambda})\mathbf{B}_{\lambda}\Big)\cdot \partial_\mathbf{r}f_{\lambda}\right]=0,
\label{CKT-kinetic-equation}
\end{equation}
where $\mathbf{\Omega}_{\lambda} =\lambda \hbar\mathbf{p}/(2|\mathbf{p}|^3)$ is the Berry curvature \cite{Berry:1984},
$\tilde{\mathbf{E}}_{\lambda} = \mathbf{E}_{\lambda}-(1/e)\partial_\mathbf{r}\epsilon_{\mathbf{p}}$, the factor
$1/[1+e(\mathbf{B}_{\lambda}\cdot\mathbf{\Omega}_{\lambda})/c]$ accounts for the correct definition of the phase-space
volume that satisfies Liouville's theorem \cite{Xiao,Duval}, and we introduced the following effective electric and magnetic fields for
fermions of a given chirality:
\begin{eqnarray}
\mathbf{E}_{\lambda}=\mathbf{E}+\lambda\mathbf{E}_{5}, \qquad \mathbf{B}_{\lambda}=\mathbf{B}+\lambda\mathbf{B}_{5}.
\label{CKT-fields}
\end{eqnarray}
In Weyl materials, the pseudoelectric field $\mathbf{E}_{5}$ can be generated by dynamical deformations of the sample
and the pseudomagnetic field $\mathbf{B}_{5}$ can be induced by a static torsion or bending
\cite{Cortijo:2016yph,Liu-Pikulin:2016,Pikulin:2016}.
The fermion energy $\epsilon_{\mathbf{p}}$ in the presence of a weak effective magnetic field $\mathbf{B}_{\lambda}$,
$\hbar|e\mathbf{B}_{\lambda}|/(c p^2) \ll 1$, is given by \cite{Son}
\begin{equation}
\epsilon_{\mathbf{p}}= v_Fp\left[1 - \frac{e}{c}(\mathbf{B}_{\lambda}\cdot \mathbf{\Omega}_{\lambda})\right],
\label{CKT-epsilon_p}
\end{equation}
where $v_F$ is the Fermi velocity, $p\equiv|\mathbf{p}|$, $e$ is an electric charge ($e<0$ for the electron), $c$ is the speed of light, and
$\hat{\mathbf{p}}=\mathbf{p}/p$. The quasiparticle velocity is defined as follows:
\begin{equation}
\mathbf{v}= \partial_\mathbf{p}\epsilon_{\mathbf{p}}
=v_F\hat{\mathbf{p}} \left[1+2\frac{e}{c} \left(\mathbf{B}_{\lambda} \cdot \mathbf{\Omega}_{\lambda}\right) \right]
- \frac{e v_F}{c}\mathbf{B}_{\lambda}\left(\hat{\mathbf{p}} \cdot \mathbf{\Omega}_{\lambda}\right).
\label{CKT-v}
\end{equation}
In equilibrium, the function $f_{\lambda}$ is given by the Fermi-Dirac distribution
\begin{equation}
f^{\rm (eq)}_{\lambda} =\frac{1}{e^{(\epsilon_{\mathbf{p}}-\mu_{\lambda})/T}+1},
\label{CKT-equilibrium-function}
\end{equation}
where $\mu_{\lambda}=\mu+\lambda\mu_5$ denotes the effective chemical potential for the left- ($\lambda=-$)
and right-handed ($\lambda=+$) fermions, $\mu$ is the electric chemical potential, $\mu_5$ is the chiral chemical
potential, and $T$ is temperature. The equilibrium distribution function for holes (antiparticles) $\bar{f}^{\rm (eq)}_{\lambda}$
is obtained by replacing $\mu_{\lambda}\to - \mu_{\lambda}$. In addition, in the chiral kinetic equation for the
hole distribution function, one should change the sign of the electric charge and the Berry curvature $\mathbf{\Omega}_{\lambda}\to-\mathbf{\Omega}_{\lambda}$.

By definition, the electric charge density consists of the left- and right-handed fermion contributions, i.e.,
$\rho=\sum_{\lambda=\pm}\rho_{\lambda}$, where
\begin{equation}
\rho_{\lambda}=\sum_{\rm p,a} e\int\frac{d^3p}{(2\pi \hbar)^3}
\left[1+\frac{e}{c}(\mathbf{B}_{\lambda}\cdot\mathbf{\Omega}_{\lambda})\right]f_{\lambda}.
\label{CKT-charge-density}
\end{equation}
The current densities of the left- and right-handed fermions are \cite{Son-Spivak,Son}
\begin{eqnarray}
\mathbf{j}_{\lambda} &=& \sum_{\rm p,a}e\int\frac{d^3p}{(2\pi \hbar)^3}\left[\mathbf{v}
+\frac{e}{c} \epsilon_{\mathbf{p}} \mathbf{B}_{\lambda} (\partial_{\mathbf{p}}\cdot \mathbf{\Omega}_{\lambda})
+\frac{e}{c}(\mathbf{v}\cdot\mathbf{\Omega}_{\lambda}) \mathbf{B}_{\lambda}
+e(\tilde{\mathbf{E}}_{\lambda}\times\mathbf{\Omega}_{\lambda})\right]\,f_{\lambda}
+\sum_{\rm p,a}e\partial_{\mathbf{r}}\times \int\frac{d^3p}{(2\pi \hbar)^3} f_{\lambda}\epsilon_{\mathbf{p}}\mathbf{\Omega}_{\lambda}.
\label{CKT-electric-current-b}
\end{eqnarray}
where $\sum_{\rm p,a}$ denotes the summation over particles and antiparticles (holes).
Therefore, the electric current density is given by $\mathbf{j}=\sum_{\lambda=\pm}\mathbf{j}_{\lambda}$.
Note that the last term in Eq.~(\ref{CKT-electric-current-b}) is the magnetization current.

By using Eqs.~(\ref{CKT-charge-density}) and (\ref{CKT-electric-current-b}) together with the Maxwell equations,
one finds that the chiral and electric currents satisfy the following continuity equations:
\begin{eqnarray}
\label{CKT-dn/dt-n5}
\partial_t\rho_5 +\partial_\mathbf{r}\cdot\mathbf{j}_5 &=& \frac{e^3}{2\pi^2 \hbar^2 c}
\Big[(\mathbf{E}\cdot\mathbf{B}) +(\mathbf{E}_{5}\cdot\mathbf{B}_{5})\Big],\\
\label{CKT-dn/dt-n}
\partial_t\rho +\partial_\mathbf{r}\cdot \mathbf{j} &=& \frac{e^3}{2\pi^2 \hbar^2 c}
\Big[(\mathbf{E}\cdot\mathbf{B}_{5}) +(\mathbf{E}_{5}\cdot\mathbf{B})\Big].
\end{eqnarray}
The first equation describes the anomalous chiral charge nonconservation \cite{ABJ} and can be understood
as a pumping of the chiral charge between the Weyl nodes of opposite chiralities. The second equation, naively,
describes the anomalous local nonconservation of the electric charge when both electromagnetic and
pseudoelectromagnetic fields are present. As emphasized in Refs.~\cite{Gorbar:2016ygi,Gorbar:2016sey},
the conservation of the electric charge in the chiral kinetic theory must be enforced locally by using the
consistent definition of the electric current \cite{Bardeen,Landsteiner:2013sja,Landsteiner:2016}:
\begin{equation}
J^{\nu} \equiv (c\rho+c\delta \rho, \mathbf{j}+\delta \mathbf{j}),
\label{consistent-4-current}
\end{equation}
where
\begin{equation}
\delta j^{\mu} =  \frac{e^3}{4\pi^2 \hbar^2 c} \epsilon^{\mu \nu \rho \lambda} A_{\nu}^5 F_{\rho \lambda}
\label{consistent-def-0}
\end{equation}
and $A_{\nu}^5=b_{\nu}+\tilde{A}^5_{\nu}$ is the axial potential, which is an observable quantity.
Indeed, while $b_0$ and $\mathbf{b}$ correspond to energy and momentum-space separations
of the Weyl nodes, respectively, $\tilde{A}_{\nu}^5$ describes strain-induced axial or, equivalently,
pseudoelectromagnetic field directly related to the deformation tensor
\cite{Zubkov:2015,Cortijo:2016yph,Cortijo:2016,Grushin-Vishwanath:2016,Pikulin:2016,Liu-Pikulin:2016}. From a physics viewpoint, the additional contributions to electric current (\ref{consistent-def-0}) capture the local changes of $\rho$ and $\mathbf{j}$ associated with the
deformations of the crystal lattice that cannot be captured  in other ways by the chiral kinetic theory of low-energy quasiparticles.

As is easy to check, the consistent electric current is nonanomalous, $\partial_{\nu} J^{\nu}=0$, and, therefore, the electric charge
is locally conserved. It is useful to rewrite the topological contribution (\ref{consistent-def-0}) explicitly in components
\begin{eqnarray}
\delta \rho &=&\frac{e^3}{2\pi^2 \hbar^2c^2}\,(\mathbf{b}\cdot\mathbf{B}),
\label{consistent-charge-density}
 \\
\delta \mathbf{j} &=&\frac{e^3}{2\pi^2 \hbar^2 c}\,b_0 \,\mathbf{B} - \frac{e^3}{2\pi^2 \hbar^2 c}\,(\mathbf{b}\times\mathbf{E}),
\label{consistent-current-density}
\end{eqnarray}
where we assumed that the field $\mathbf{B}_{5}$ is weak or, in other words, that $\tilde{\mathbf{A}}^5$ is negligible compared to the chiral
shift $\mathbf{b}$.

The first term in $\delta \mathbf{j}$  at $b_0=-\mu_5/e$ leads to the cancellation of the CME current in the equilibrium state
\cite{Landsteiner:2016} as is required for solids \cite{Franz,Basar}. The second term in Eq.~(\ref{consistent-current-density})
describes the anomalous Hall effect in Weyl materials \cite{Burkov:2011ene,Grushin-AHE, Goswami} in the framework of the semiclassical
kinetic theory \cite{Gorbar:2016ygi}.

\section{Coefficients $g_n$}
\label{sec:App-g}

In this appendix, we present the explicit results for the Fourier coefficients $g_n$ defined by Eq.~(\ref{collective-B-gn-def}).
The integral representation for the Fourier coefficients is given by
\begin{eqnarray}
g_n  &=&  -\frac{i}{2\pi(a_1+n)} \int_{0}^{2\pi} d\tau \,e^{i a_2 \sin\tau-in \tau}
\Bigg\{ a_4 -\frac{a_5}{2p^2} E_{\perp} k_{\perp}k_{\parallel} p_{\perp}^2 \sin{\phi_E} \nonumber\\
&+&e^{i \phi} \Big[ \frac{a_3}{2}e^{-i\phi_E} +\frac{a_5}{2p^2} E_{\perp}k_{\perp}^2p_{\perp}p_{\parallel} \sin{\phi_E} +i\frac{a_5}{2p^2}E_{\parallel}k_{\perp}k_{\parallel}p_{\perp}p_{\parallel} -i\frac{a_5}{2p^2}E_{\perp}k_{\parallel}^2p_{\perp}p_{\parallel}e^{-i\phi_E} \Big] \nonumber\\
&+&e^{-i \phi} \Big[ \frac{a_3}{2}e^{i\phi_E} +\frac{a_5}{2p^2} E_{\perp}k_{\perp}^2p_{\perp}p_{\parallel} \sin{\phi_E} -i\frac{a_5}{2p^2}E_{\parallel}k_{\perp}k_{\parallel}p_{\perp}p_{\parallel}  +i\frac{a_5}{2p^2}E_{\perp}k_{\parallel}^2p_{\perp}p_{\parallel}e^{i\phi_E}  \Big] \nonumber\\
&+& ie^{2i\phi} \frac{a_5 p_{\perp}^2}{4p^2}\Big[ E_{\parallel}k_{\perp}^2 -e^{-i\phi_E}E_{\perp}k_{\perp}k_{\parallel}\Big]
-ie^{-2i\phi} \frac{a_5 p_{\perp}^2}{4p^2}\Big[ E_{\parallel}k_{\perp}^2 -e^{i\phi_E}E_{\perp}k_{\perp}k_{\parallel}\Big]
\Bigg\},
\end{eqnarray}
where coefficients $a_i$ with $i=\overline{1,5}$ are given by Eq.~(\ref{collective-Q-ee}). After performing the integration over $\tau$, we derive
\begin{eqnarray}
g_n  &=& -\frac{i}{a_1+n} \Bigg\{J_n(a_2)\Big[a_4 - \frac{a_5}{2p^2} E_{\perp} k_{\perp}k_{\parallel} p_{\perp}^2 \sin{\phi_E}\Big] \nonumber\\
&+& J_{n-1}(a_2)\Big[ \frac{a_3}{2}e^{-i\phi_E} +\frac{a_5}{2p^2} E_{\perp}k_{\perp}^2p_{\perp}p_{\parallel} \sin{\phi_E} +i\frac{a_5}{2p^2}E_{\parallel}k_{\perp}k_{\parallel}p_{\perp}p_{\parallel}  -i\frac{a_5}{2p^2}E_{\perp}k_{\parallel}^2p_{\perp}p_{\parallel}e^{-i\phi_E} \Big]\nonumber\\
&+& J_{n+1}(a_2)\Big[ \frac{a_3}{2}e^{i\phi_E} +\frac{a_5}{2p^2} E_{\perp}k_{\perp}^2p_{\perp}p_{\parallel} \sin{\phi_E} -i\frac{a_5}{2p^2}E_{\parallel}k_{\perp}k_{\parallel}p_{\perp}p_{\parallel} +i\frac{a_5}{2p^2}E_{\perp}k_{\parallel}^2p_{\perp}p_{\parallel}e^{i\phi_E} \Big]\nonumber\\
&+& i J_{n-2}(a_2) \frac{a_5 p_{\perp}^2}{4p^2}\Big[ E_{\parallel}k_{\perp}^2 -e^{-i\phi_E}E_{\perp}k_{\perp}k_{\parallel}\Big]
-iJ_{n+2}(a_2) \frac{a_5 p_{\perp}^2}{4p^2}\Big[ E_{\parallel}k_{\perp}^2 -e^{i\phi_E}E_{\perp}k_{\perp}k_{\parallel}\Big]
\Bigg\}.
\label{helicon-gn-1}
\end{eqnarray}
Here we used the table integral
\begin{equation}
J_{n}(x) =\frac{1}{2\pi} \int_{0}^{2\pi} e^{i (n \theta- x \sin\theta)} d\theta,
\end{equation}
and the following identities for the Bessel functions: $J_{-n}(x) =(-1)^{n}J_{n}(x)$ and $J_{n}(-x) =(-1)^{n}J_{n}(x)$
(see formulas 8.411.1 and 8.404.2 in Ref.~\cite{Gradshtein}).

\section{Polarization vector}
\label{sec:App-P}

In this appendix, we present the details of calculation of the polarization vector $\mathbf{P}$ in the limit of small $\omega$.
The polarization vector (\ref{collective-B-polarization}) at $n=0,\pm1$ reads as
\begin{eqnarray}
\mathbf{P} &\simeq& \sum_{\lambda=\pm} \sum_{\rm p,a} \frac{ie}{\omega}\int\frac{d^3p}{(2\pi \hbar)^3} \left[ e(\tilde{\mathbf{E}}\times\mathbf{\Omega}_{\lambda})
+\frac{e}{\omega}(\mathbf{v}\cdot\mathbf{\Omega}_{\lambda})\left( \mathbf{k} \times \mathbf{E} \right) +\frac{e}{c}(\mathbf{\delta v}
\cdot\mathbf{\Omega}_{\lambda})\mathbf{B}_{0,\lambda}\right] f_{\lambda}^{\rm (eq)} \nonumber\\
&+&\sum_{\lambda=\pm}\sum_{\rm p,a} \frac{\lambda e^2 \hbar v_F}{2\omega^2}\int\frac{d^3p}{(2\pi \hbar)^3} \frac{1}{p} f_{\lambda}^{\rm (eq)} [\mathbf{k}\times
\mathbf{\Omega}_{\lambda}] \left(\hat{\mathbf{p}}\cdot[\mathbf{k}\times\mathbf{E}]\right)  -i \frac{e^3}{2\pi^2 \omega c \hbar^2}(\mathbf{b}\times \mathbf{E}) + i \frac{e^3 b_0}{2\pi^2\omega^2 \hbar^2} (\mathbf{k}\times \mathbf{E})\nonumber\\
&+&\sum_{\lambda=\pm} \sum_{n=-1}^1\sum_{\rm p,a}\frac{ie}{\omega}\int\frac{d^3p}{(2\pi \hbar)^3} v_F\hat{\mathbf{p}}\left[1+2\frac{e}{c}(\mathbf{B}_{0,\lambda}\cdot\mathbf{\Omega}_{\lambda}) \right] g_n e^{in\phi} +O(B_{0,\lambda}^2),
\label{collective-B-polarization-1}
\end{eqnarray}
where we neglected terms of order $O(B_{0,\lambda}^2)$ and terms suppressed by powers of the
wave vector, such as $k f^{(1)}_{\lambda, n}$. Also, $\delta \mathbf{v}$ is defined in Eq.~(\ref{collective-B-delta-v}). Using the expansion of the equilibrium distribution function (\ref{collective-B-f-series}), the first two terms in
Eq.~(\ref{collective-B-polarization-1}) can be represented in the following form (we dropped the summation over particles and
antiparticles):
\begin{eqnarray}
\label{collective-B-X1-a}
 \mathbf{X}_1^{\rm  (a)} &=& \frac{ie^2}{\omega}\int\frac{d^3p}{(2\pi \hbar)^3} (\mathbf{E}\times\mathbf{\Omega}_{\lambda}) f_{\lambda}^{\rm (eq)}
\simeq -i\frac{e^3\hbar^2 v_F (\mathbf{E}\times\mathbf{B}_{0,\lambda})}{12 c \omega}\int\frac{d^3p}{(2\pi \hbar)^3}\frac{1}{p^3}
\frac{\partial f^{(0)}_{\lambda}}{\partial \epsilon_{\mathbf{p}}} +O\left(B_{0,\lambda}^3\right), \\
\label{collective-B-X1-b}
 \mathbf{X}_1^{\rm  (b)} &=& -\frac{e^2}{\omega}\int\frac{d^3p}{(2\pi \hbar)^3}  \frac{\lambda \hbar v_F }{2\omega p} (\mathbf{E} \cdot[\mathbf{k}
 \times\hat{\mathbf{p}}])(\mathbf{k}\times\mathbf{\Omega}_{\lambda})f_{\lambda}^{\rm (eq)} \simeq -\frac{v_Fe^2}{24\pi^2 \omega^2 \hbar} \int\frac{dp}{p} f^{(0)}_{\lambda} \left[\mathbf{k}\times(\mathbf{k}\times\mathbf{E})\right] +O\left(B_{0,\lambda}^2\right),\\
\label{collective-B-X1-c}
 \mathbf{X}_1^{\rm  (c)} &=& \frac{ie^2}{\omega}\int\frac{d^3p}{(2\pi \hbar)^3} \frac{(\mathbf{v}\cdot\mathbf{\Omega}_{\lambda})}{\omega} \left( \mathbf{k} \times
 \mathbf{E} \right)  f_{\lambda}^{\rm (eq)} \simeq  i\left( \mathbf{k} \times \mathbf{E} \right)  \frac{\lambda e^2 T}{4\pi^2\hbar^2 \omega^2} \ln
 \left(1+e^{\mu_\lambda/T}\right) +O\left(B_{0,\lambda}^2\right),\\
\label{collective-B-X1-d}
\mathbf{X}_1^{\rm  (d)} &=& \frac{ie^2}{\omega} \int\frac{d^3p}{(2\pi \hbar)^3} \frac{(\mathbf{\delta v}\cdot\mathbf{\Omega}_{\lambda})\mathbf{B}_{0,\lambda}}{c}
f_{\lambda}^{\rm (eq)} \simeq i\int\frac{d^3p}{(2\pi \hbar)^3} f^{(0)}_{\lambda} \frac{e^3 \hbar^2 v_F \mathbf{B}_{0,\lambda}}
{4c\omega^2 p^4} \left(\hat{\mathbf{p}}\cdot[\mathbf{k}\times\mathbf{E}]\right) =O\left(B_{0,\lambda}^2\right),\\
\label{collective-B-X1-e}
\mathbf{X}_1^{\rm  (e)}&=&-\mathbf{X}_1^{\rm  (b)},
\end{eqnarray}
where $f^{(0)}_{\lambda} =1/[e^{(\epsilon^{(0)}_{\mathbf{p}}-\mu_{\lambda})/T}+1]$ is the equilibrium function
at $\mathbf{B}_{0,\lambda}=0$ and $\epsilon^{(0)}_{\mathbf{p}}=v_Fp$.
By adding the contribution of holes (antiparticles), we obtain
\begin{eqnarray}
\label{collective-B-X1}
\mathbf{X}_1 &=& \sum_{\rm p, a} \left(\mathbf{X}_1^{\rm (a)}+\mathbf{X}_1^{\rm (b)}+\mathbf{X}_1^{\rm (c)}+\mathbf{X}_1^{\rm  (e)}\right)=
-i\frac{e^3\hbar^2v_F(\mathbf{E}\times\mathbf{B}_{0,\lambda})}{12c\omega} \int\frac{d^3p}{(2\pi \hbar)^3}\frac{1}{p^3}\left(
\frac{\partial f^{(0)}_{\lambda}}{\partial \epsilon_{\mathbf{p}}} -\frac{\partial \bar{f}^{(0)}_{\lambda}}{\partial \epsilon_{\mathbf{p}}} \right)  \nonumber\\
&+& \frac{i\lambda T e^2}{4\pi^2 \hbar^2 \omega^2} \left( \mathbf{k} \times \mathbf{E}\right) \left[\ln
\left(1+e^{\mu_\lambda/T}\right) -\ln\left(1+e^{-\mu_\lambda/T}\right)\right] =i\frac{e^3v_F(\mathbf{E}\times\mathbf{B}_{0,\lambda})}{24\pi^2 \hbar \omega cT} F\left(\frac{\mu_\lambda}{T}\right)
 + \frac{i\lambda \mu_\lambda e^2}{4\pi^2 \hbar^2 \omega^2} \left( \mathbf{k} \times \mathbf{E} \right) +O(B_{0,\lambda}^2),\nonumber\\
\end{eqnarray}
where the replacements $e\to-e$, $\mu_{\lambda}\to -\mu_{\lambda}$, and $\lambda \to -\lambda$ in the Berry curvature
$\mathbf{\Omega}_{\lambda}$ were made for holes (antiparticles). Further, the function
\begin{eqnarray}
\label{collective-B-F1-def}
F\left(\nu_\lambda \right) \equiv -T\int\frac{dp}{p}\left( \frac{\partial f^{(0)}_{\lambda}}{\partial \epsilon_{\mathbf{p}}}
-\frac{\partial \bar{f}^{(0)}_{\lambda}}{\partial \epsilon_{\mathbf{p}}} \right)
\end{eqnarray}
can be easily computed numerically.
Thus, Eq.~(\ref{collective-B-polarization-1}) can be rewritten as
\begin{eqnarray}
\mathbf{P}
&\simeq& \sum_{\lambda=\pm} i\frac{e^3v_F(\mathbf{E}\times\mathbf{B}_{0,\lambda})}{24\pi^2 \hbar \omega cT} F\left(\frac{\mu_\lambda}{T}\right) -i \frac{e^3}{2\pi^2 \omega c \hbar^2}(\mathbf{b}\times \mathbf{E}) +i \frac{e^2 (eb_0+\mu_5)}{2\pi^2\omega^2 \hbar^2} (\mathbf{k}\times \mathbf{E}) \nonumber\\
&+&\sum_{\lambda=\pm} \sum_{n=\pm}\sum_{\rm p,a}\frac{\pi e^2v_F^2}{2\omega}\int_0^{\infty}\frac{dp}{(2\pi \hbar)^3} \int_{-1}^{1} d\cos{\theta}\, p_{\perp}^2 \left(1 +\frac{3\lambda \hbar e p_{\parallel} B_{0,\lambda}}{2cp^3}\right) \frac{\left(\mathbf{E}-\hat{\mathbf{z}}(\hat{\mathbf{z}}\cdot\mathbf{E})\right)-in(\hat{\mathbf{z}}\times\mathbf{E})}{\omega+n\Omega_c} \frac{\partial f_{\lambda}^{\rm (eq)}}{\partial \epsilon_{\mathbf{p}}}\nonumber\\
&+&\sum_{\lambda=\pm} \sum_{\rm p,a}\frac{2\pi e^2v_F^2}{\omega^2}\int_0^{\infty}\frac{dp}{(2\pi \hbar)^3} \int_{-1}^{1} d\cos{\theta} \, \, \hat{\mathbf{z}}(\mathbf{E}\cdot\hat{\mathbf{z}}) \,p_{\parallel}^2\left(1 +\frac{3\lambda \hbar e p_{\parallel} B_{0,\lambda}}{2cp^3}\right)
\frac{\partial f_{\lambda}^{\rm (eq)}}{\partial \epsilon_{\mathbf{p}}} + O(B_{0,\lambda}^2).
\label{collective-B-polarization-2}
\end{eqnarray}
Let us consider the case of small frequencies $\omega\ll\Omega_c|_{p=p^{*}}$, where $\Omega_c$ is given in Eq.~(\ref{collective-B-Omegac}) and
$p^{*} \sim \sqrt{\mu_5^2 +\mu^2+\pi^2T^2}/v_F$ is a characteristic momentum, relevant for helicons.
In the zero order in $\omega/\Omega_c$, the fourth term of the above equation equals
\begin{eqnarray}
\sum_{n=\pm}\mathbf{P}_n^{(0)} &\simeq&\sum_{\lambda=\pm} \sum_{n=\pm}\sum_{\rm p,a}  \left[n\left(\mathbf{E}-\hat{\mathbf{z}}(\hat{\mathbf{z}}\cdot\mathbf{E})\right)-i(\hat{\mathbf{z}}\times\mathbf{E})\right] \frac{ev_F c}{12\pi^2\hbar^3B_{0,\lambda}\omega} \int_0^{\infty} dp\,p^3 \frac{\partial f^{(0)}_{\lambda}}{\partial \epsilon_{\mathbf{p}}} +O(B_{0,\lambda}^2)\nonumber\\
&=&\sum_{\lambda=\pm} \sum_{n=\pm}\sum_{\rm p,a} \left[n\left(\mathbf{E}-\hat{\mathbf{z}}(\hat{\mathbf{z}}\cdot\mathbf{E})\right)-i(\hat{\mathbf{z}}\times\mathbf{E})\right] \frac{ecT^3}{2\pi^2 \hbar^3 B_{0,\lambda}v_F^3\omega} \mbox{Li}_3\left(-e^{\mu_{\lambda}/T}\right) +O(B_{0,\lambda}^2)\nonumber\\
&=&i\sum_{\lambda=\pm} [\mathbf{E}\times\hat{\mathbf{z}}] \frac{e c \mu_{\lambda}}{6\pi^2\hbar^3B_{0,\lambda}v_F^3 \omega} \left(\mu_{\lambda}^2+\pi^2T^2\right)+O(B_{0,\lambda}^2).
\label{collective-B-polarization-delta0}
\end{eqnarray}

To the leading order in small $\omega/\Omega_c$, we have
\begin{eqnarray}
\sum_{n=\pm}\mathbf{P}_n^{(1)} &\simeq&-\sum_{\lambda=\pm} \sum_{n=\pm}\sum_{\rm p,a} \left[\left(\mathbf{E}-\hat{\mathbf{z}}(\hat{\mathbf{z}}\cdot\mathbf{E})\right)-in(\hat{\mathbf{z}}\times\mathbf{E})\right] \frac{c^2}{12\pi^2\hbar^3B_{0,\lambda}^2} \int_0^{\infty} dp \, p^4 \frac{\partial f^{(0)}_{\lambda}}{\partial \epsilon_{\mathbf{p}}} +O(B_{0,\lambda}^2)\nonumber\\
&=&-\sum_{\lambda=\pm} \sum_{n=\pm}\sum_{\rm p,a} \left[\left(\mathbf{E}-\hat{\mathbf{z}}(\hat{\mathbf{z}}\cdot\mathbf{E})\right)-in(\hat{\mathbf{z}}\times\mathbf{E})\right]
\frac{2c^2T^4}{\pi^2\hbar^3B_{0,\lambda}^2v_F^5} \mbox{Li}_4\left(-e^{\mu_{\lambda}/T}\right)+O(B_{0,\lambda}^2)\nonumber\\
&=&\sum_{\lambda=\pm} \left[\mathbf{E}-\hat{\mathbf{z}}(\mathbf{E}\cdot\hat{\mathbf{z}})\right] \frac{c^2}{6\pi^2\hbar^3B_{0,\lambda}^2v_F^5} \left(\mu_{\lambda}^4 + 2\pi^2 \mu_{\lambda}^2T^2+\frac{7\pi^4T^4}{15}\right)+O(B_{0,\lambda}^2).
\label{collective-B-polarization-delta1}
\end{eqnarray}
The contribution of $g_0$ to the polarization vector is given by the last term in Eq.~(\ref{collective-B-polarization-1}).
Its explicit form reads
\begin{eqnarray}
\mathbf{P}_0 &\simeq& \hat{\mathbf{z}}(\mathbf{E}\cdot\hat{\mathbf{z}})\sum_{\lambda=\pm} \sum_{\rm p,a} \int_0^{\infty} dp \frac{e^2 v_F^2}{6\pi^2 \hbar^3 \omega^2} p^2\frac{\partial f_{\lambda}^{(0)}}{\partial \epsilon_{\mathbf{p}}}+O(B_{0,\lambda}^2) = \hat{\mathbf{z}}(\mathbf{E}\cdot\hat{\mathbf{z}})\sum_{\lambda=\pm} \sum_{\rm p,a} \int_0^{\infty} dp \frac{e^2 T^2}{3\pi^2 \hbar^3 v_F\omega^2}  \mbox{Li}_2\left(-e^{\mu_{\lambda}/T}\right)  \nonumber\\
&+&O(B_{0,\lambda}^2)=-\hat{\mathbf{z}}(\mathbf{E}\cdot\hat{\mathbf{z}})\sum_{\lambda=\pm} \frac{e^2}{6\pi^2\hbar^3v_F \omega^2} \left(\mu_{\lambda}^2 +\frac{\pi^2T^2}{3}\right)+O(B_{0,\lambda}^2).
\label{collective-B-polarization-s=0}
\end{eqnarray}
In the derivation above, we used the following table integral:
\begin{eqnarray}
\int\frac{d^3p}{(2\pi)^3} p^{n-2} \frac{\partial f^{(0)}_{\lambda}}{\partial \epsilon_{\mathbf{p}}}
&=& \frac{T^{n} \Gamma(n+1) }{2\pi^2 v_F^{n+1}}  \mbox{Li}_{n}\left(-e^{\mu_{\lambda}/T}\right),
\qquad n\geq 0,
\label{integral-3b}
\end{eqnarray}
where $\mbox{Li}_{n}(x)$ is the polylogarithm function (see formula 1.1.14 in Ref.~\cite{Erdelyi:Vol1}
where $\mathrm{Li}_n(x) \equiv \mathrm{F}(x, n)$). We also used the following identities for the
polylogarithm functions:
\begin{eqnarray}
\mbox{Li}_{2} (-e^{x}) +\mbox{Li}_{2} (-e^{-x})   &=& -\frac{x^2}{2}-\frac{\pi^2}{6},
\\
\mbox{Li}_{3} (-e^{x}) - \mbox{Li}_{3} (-e^{-x})   &=&  -\frac{x^3}{6}-\frac{\pi^2 x}{6},\\
\mbox{Li}_{4} (-e^{x}) + \mbox{Li}_{4} (-e^{-x})  &=& -\frac{x^4}{24}-\frac{\pi^2 x^2}{12}-\frac{7\pi^4}{360} .
\end{eqnarray}

\end{document}